\documentclass{aa}
\usepackage{txfonts}
\usepackage{natbib}
\bibpunct{(}{)}{;}{a}{}{,}

\usepackage{ifthen}
\ifx\pdftexversion\undefined
\usepackage[dvips]{graphicx}
\usepackage{psfig}

\else
\usepackage[pdftex]{graphicx}
\usepackage{epstopdf}
\usepackage[pdftex,pdftitle={Magnetic turbulence in cool cores of galaxy clusters},pdfauthor={Torsten En{\ss}lin},a4paper]{hyperref}
\fi

\begin{document}

\def\eps{\varepsilon}
\def\aap{A\&A}
\def\araa{Ann. Rev. of Astr. \& Astrophys.}
\def\apj{ApJ}
\def\apjl{ApJL}
\def\mnras{MNRAS}
\def\aj{AJ}
\def\nat{Nature}
\def\aaps{A\&A Supp.}
\def\e{{\rm e}}
\def\p{{\rm p}}
\def\me{m_\e}
\def\mp{m_\p}
\def\CR{{\rm CR}}
\def\th{{\rm th}}
\def\lesssim{\mathrel{\hbox{\rlap{\hbox{\lower4pt\hbox{$\sim$}}}\hbox{$<$}}}}
\def\gtrsim{\mathrel{\hbox{\rlap{\hbox{\lower4pt\hbox{$\sim$}}}\hbox{$>$}}}}
\def\vx{\vec{x}}
\def\vr{\vec{r}}
\def\fl{{\rm fl}}
\def\etal{{\it et al.}}
\def\eg{{\it eg.}}
\def\B{{\mathcal B}}

\def\RM{\textit{RM }}

\def\cc{\mathrm{cc}}
\def\Mcc{M_\mathrm{cc}}
\def\Tcc{T_\mathrm{cc}}
\def\Rcc{r_\mathrm{cc}}
\def\Vcc{V_\mathrm{cc}}
\def\ncc{n_\mathrm{cc}}
\def\Lcc{L_\mathrm{cc}}
\def\Pcc{P_\mathrm{cc}}
\def\LX{\Lambda_\mathrm{X}}
\def\L0{\Lambda_{0}}

\def\lB{\lambda_B}
\def\lT{\lambda_\mathrm{T}}
\def\epsT{\eps_\mathrm{T}}
\def\etaT{\eta_\mathrm{T}}
\def\fT{f_\mathrm{T}}
\def\vT{\mathrm{v}_\mathrm{T}}
\def\Rc{R_\mathrm{c}}

\def\del#1{{}}

  
\title{Magnetic turbulence in cool cores of galaxy clusters}
\titlerunning{Magnetic turbulence in cool cores}
\author{Torsten A. En{\ss}lin\inst{1} \and Corina Vogt\inst{1,}\inst{2}}
\authorrunning{T. A. En{\ss}lin \& C. Vogt}
\institute{Max-Planck-Institut f\"{u}r Astrophysik,
Karl-Schwarzschild-Str.1, Postfach 1317, 85741 Garching, 
Germany
	\and 
ASTRON, P.O.Box 2, 7990 AA Dwingeloo, The Netherlands} 
\date{\today}

\abstract{We argue that the recently reported Kolmogorov-like magnetic
turbulence spectrum in the cool core of the Hydra~A galaxy cluster can be
understood by kinetic energy injection by active galaxies that drives a
turbulent non-helical magnetic dynamo into its saturated state. Although
dramatic differences exist between small-scale dynamo scenarios, their
saturated state is expected to be similar, as we show for three scenarios: the
{\it flux rope dynamo}, the {\it fluctuation dynamo}, and the {\it explosive
dynamo}. Based on those scenarios, we develop an analytical model of the
hydrodynamic and magnetic turbulence in cool cores. The model implies magnetic
field strengths that fit well with Faraday rotation measurements and minimum
energy estimates for the sample of cool core clusters having such data
available. Predictions for magnetic fields in clusters for which the
appropriate observational information is still missing, and for yet unobserved
quantities like the hydrodynamical turbulence velocity and characteristic
length-scale are provided. The underlying dynamo models suggest magnetic
intermittency and possibly a large-scale hydrodynamic viscosity. We conclude
that the success of the model to explain the field strength in cool core
clusters indicates that in general cluster magnetic fields directly reflect
hydrodynamical turbulence, also in clusters without cool cores.
 } \maketitle

\section{Introduction}
\subsection{Magnetic fields in cool cores}

Galaxy clusters contain magnetised plasma on cluster scales, as
radio-synchrotron emission of relativistic electrons reveals in form
of the so-called {\it cluster radio halos}. The origin, the strength
and geometry of these magnetic fields is still a mystery.

With the emerging of the first theories of magnetic dynamos
\citep{Batchelor1950, kazantsev67, 1990alch.book.....Z}, it was proposed that
the cluster magnetic fields are the product of turbulence acting on a seed
magnetic field. The seed field could be a remnant of non-equilibrium processes
in the early Universe, or due to some weak magnetisation caused by galactic
outflows in the form of winds and radio plasma \citep[for a review
see][]{2002RvMP...74..775W}.

Initially, wakes of the weakly sonic galaxy motions were thought to be the main
driver of the intra-cluster
turbulence\footnote{e.g. \citet{1980ApJ...241..925J, 1981A&A....93..407R,
1989MNRAS.241....1R, 1991ApJ...380..344G, 1992ApJ...386..464D}}. However, it
was realised using numerical simulations of large-scale structure formation
that the violent mergers of galaxy clusters are a much more powerful source of
turbulence \citep[e.g.][]{1993MNRAS.263...31T}. In numerical simulations of
magnetic field amplification, the merger-driven turbulence of galaxy clusters
seems to be able to reproduce the typical field values of
clusters\footnote{\citet{1999A&A...348..351D, 1999ApJ...518..594R,
2002A&A...387..383D}}.

It is therefore surprising that the strongest magnetic fields seem to be
located in the centres of clusters which are dynamically the most relaxed,
since the last major merger was long ago. These {\it cooling flow} (now {\it
cool core}) clusters had time to develop a cool, dense central region due to
the cooling instability of optically thin X-ray emission of cooling gas. The
magnetic fields reported for these cooling flow regions inferred by Faraday
rotation studies were extraordinarily strong (up to $50\, \mu$G) compared to
the few $\mu$G fields reported for non-cooling flow clusters \citep[for recent
reviews see][]{2002ARA&A..40..319C, 2004IJMPD..13.1549G}. It was speculated
that the strong fields could be a result of compression in the cooling flow
\citep{1990ApJ...348...73S}.

However, Chandra and XMM observations revealed that the standard cooling flow
picture, in which unheated gas cools down to neutral gas temperatures, must be
incorrect, since the expected amount of line emission of cold ($< 0.3$ keV) gas
or the expected number of stars formed from the condensing gas was not
detected\footnote{\citet{1989ApJ...338...48H, 1991A&ARv...2..191F,
1995A&A...297...13H, 1995MNRAS.276..947A, 1997MNRAS.284L...1J,
1997ApJ...478..516S, 1998AJ....116..623O, 1999MNRAS.306..857C,
2000ApJ...545..670D, 2001MNRAS.328..762E, 2001ApJ...560..187O,
2003A&A...412..657S, 2003ApJ...594L..13E}}.  Therefore a heat source has to be
present which balances the cooling of the coldest gas. Since the theoretical
scenario associated with the term {\it cooling flow} has been ruled out
recently, the only observationally motivated term {\it cool core} is used in
the following.

The energy losses of the cool core have to be balanced by a similar energy
injection. There have been two main proposals\footnote{There are also other
mechanisms discussed in the literature, e.g. the influence of a cosmic ray
population \citep{2004ApJ...616..169C, 2005ApJ...620..191C}.}  for the heat
source in cool cores:

\begin{enumerate}
\item[(i)] Thermal conductivity, which is close to Spitzer's estimate, and therefore
  not suppressed by magnetic fields. This would allow the inward
  transport of heat from the hotter environmental intra-cluster medium
  (ICM)\footnote{\citet{2001ApJ...554..561M, 2001ApJ...562L.129N,
  2002MNRAS.335L...7V, 2002ApJ...581..223R, 2003ApJ...589L..77C,
  2004A&A...422..445S, 2004MNRAS.347.1130V, 2004MNRAS.350.1015S,
  2004MNRAS.351..423J, 2004ApJ...602..170C, 2004ApJ...606L..97D}}.
\item[(ii)] Dissipation of mechanical energy released by the expansion and
  buoyant motion of radio bubbles inflated by the radio galaxies which are
  typically found in the centre of a cool core of a galaxy
  cluster\footnote{\citet{2001ApJ...554..261C, 2001MNRAS.325..676B,
  2001MNRAS.328.1091Q, 2002MNRAS.331..545B, 2004ApJ...616..169C, 2004ApJ...617..896H,
  2005ApJ...622..205D}. There have been also observations of hot gas bubbles
  \citep{2004JKAS...37..381M}, which may also contribute to the cool core
  heating in a very similar way as the radio bubbles
  \citep{2005ApJ...622..847S}}.
\end{enumerate}

The thermal conduction scenario (i) as the only heat injection mechanism into
cool cores faces severe problems in terms of fine-tuning the required energy
injection, and explaining the existence of cold gas clouds which need
sufficient insulation from the keV gas \citep{2003MNRAS.342..463S,
2004MNRAS.349.1509N}. Too strong conduction would erase the cool core, but too
weak conduction cannot prevent the cooling catastrophe. Therefore a
conductively heated cool core should be unstable.

The scenario (ii) in which the central radio galaxy balances the radiative
energy losses of the cool core provides fine tuning in the form of a
self-adapting feedback mechanism: If the radio galaxy activity is triggered by
cold gas condensing out of the cool core onto the central galaxy, the galaxy
activity increases until it disrupts further accretion
\citep{2001ApJ...554..261C}.

In this work we assume scenario (ii), not only since it is -- at least
in our view -- theoretically more compelling, but also because it
predicts a certain level of hydrodynamical turbulence, which can be
compared to the level required to explain the cool core magnetic
fields by magnetic dynamo theory.

There have been some reports on observed signatures of turbulent flows in cool
cores of galaxy clusters, and the scenario (ii) investigated here seems to
become widely accepted, at least as a working hypothesis
\citep[e.g.][]{1990MNRAS.242..120L, 2001ApJ...554..261C, 2002A&A...382..804B,
2002MNRAS.332..729C, 2004MNRAS.347...29C}.

\subsection{Observations of cluster magnetic fields}
\label{sec:magobs}
Magnetic fields in normal clusters and cool core clusters have revealed their
existence by the diffuse radio halo emission in many clusters and radio
mini-halos in cool core regions. Furthermore, the Faraday rotation of linearly
polarised radio emission traversing the intra-cluster medium independently
proves the existence of intra-cluster magnetic fields.  If the Faraday active
medium is external to the source of the polarised emission, one expects the
change in polarisation angle to be proportional to the squared wavelength. The
proportionality factor is called the rotation measure (\textit{RM}). This
quantity can be evaluated in terms of the line of sight integral over the
product of the electron density and the magnetic field component along the line
of sight.

Magnetic fields in non-cool core clusters of galaxies detected through Faraday
rotation measurements are of the order of a few
$\mu$G. \citet{1991ApJ...379...80K} measured field strengths of about 2 $\mu$G
on scales of 10 kpc in a statistical sample of point sources observed through
the Coma cluster. \citet{2001ApJ...547L.111C}, \citet{2004JKAS...37..337C}, and
\citet{2004astro.ph.11045J} derive similar field strengths of several $\mu$G in
various samples of point sources observed within and through various low
redshift clusters in the northern and southern hemisphere. The analysis of $RM$
maps of extended radio sources have lead to the same conclusion
\citep[e.g.][]{1995A&A...302..680F, 1999A&A...344..472F, 2001MNRAS.326....2T,
2001A&A...379..807G, 2002ApJ...567..202E, 2003A&A...412..373V}.

However, the analysis of Faraday rotation measurements of extended
radio sources in the centre of cool core clusters reveal
higher magnetic field strengths. Such an analysis has been done for the
Centaurus cluster by \citet{2002MNRAS.334..769T}, for A1958 (better
known as 3C295) by \citet{1991AJ....101.1623P}, for A1795 by
\citet{1993AJ....105..778G}, for Cygnus A by
\citet{1987ApJ...316..611D} and for Hydra A by
\citet{1993ApJ...416..554T}. From these \RM measurements, magnetic
field strengths of 10 to 40 $\mu$G have been reported for the cores of
these cool core clusters on scales of 3--5 kpc.

\begin{figure}[t]
\begin{center}
\resizebox{\hsize}{!}{\includegraphics{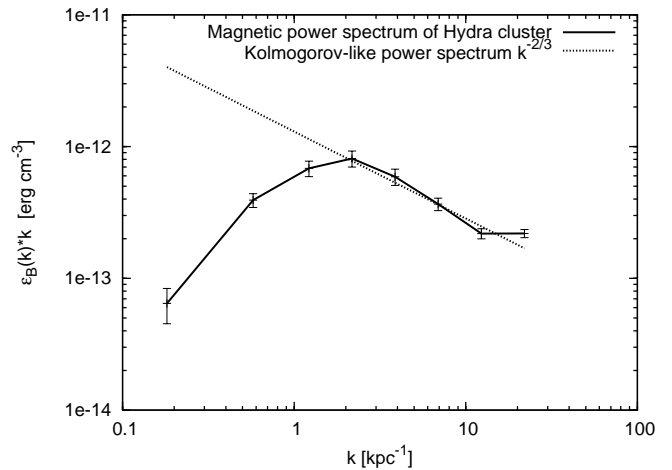}}
\end{center}
\vspace{-0.0cm}
\caption[]{Magnetic turbulence in the centre of the cool core cluster Hydra~A
as derived from the Faraday rotation map of the northern radio lobe of Hydra~A
by \citet{2005A&A...434...67V}. An angle of $45^\circ$ was assumed in this
figure between the line of sight and the approaching north-lobe. Variation of
this angle changes the overall normalisation of the spectrum, but not its
shape. The right-most data point is likely to be contaminated by observational
noise in the Faraday map. A central root-mean-square magnetic field strength of
$B_\mathrm{rms}= 7.3\pm 0.2 \pm 2 \, \mu$G and a magnetic autocorrelation
length of $\lambda_B = 2.8 \pm 0.2 \pm 0.5\, $kpc was derived by
\citet{2005A&A...434...67V}. The first errors are the statistical uncertainties
due to the limited statistics, whereas the systematic error reflects the
uncertainties in the geometry of the radio source and the Faraday screen.}
\label{fig:BturbHydra} 
\end{figure}

These rather large field values for cool core clusters have been
revised in the case of Hydra~A by a recent analysis of the
observational data. A high quality Faraday rotation map of the north
lobe of Hydra~A produced by the novel PACERMAN algorithm
\citep{2005MNRAS.358..726D, 2005MNRAS.358..732V} based on the data of
\citet{1993ApJ...416..554T} was analysed by
\citet{2005A&A...434...67V}. They used a maximum likelihood estimator
for the derivation of the magnetic power spectra, based on the theory
of turbulent Faraday screens \citep{2003A&A...401..835E,
2003A&A...412..373V}, and also using the most up-to date gas density
profile of the cool core, which turned out to make a crucial
difference. Thereby, a magnetic field strength of $7\pm 2\mu$G was
found in the centre of the cool core region of the Hydra~A cluster,
which is still a significantly larger field than reported for non-cool
core clusters. \citet{2005A&A...434...67V} measured the detailed
magnetic power spectrum from the Hydra~A dataset, which revealed a
Kolmogorov-type spectrum on small scales indicating turbulence (see
Fig.~\ref{fig:BturbHydra}).

It has been debated whether the magnetic fields seen by the Faraday effect
exist on cluster scales in the ICM, or in a mixing layer around the radio
plasma which emits the polarised emission \citep{1990ApJ...357..373B,
2003ApJ...588..143R}. However, there is no valid indication of a source local
Faraday effect in the discussed cases \citep{2003ApJ...597..870E}, and the
Faraday rotation signal excess of radio sources behind clusters compared to a
field control sample strongly supports the existence of strong magnetic fields
in the wider ICM \citep{2001ApJ...547L.111C, 2004astro.ph.11045J,
2004JKAS...37..337C}.

The morphological structure of magnetic fields in galaxy clusters is difficult
to obtain due to the projection effects in radio observations. However, there
are situations in which the magnetic fields are illuminated in only sheet-like
sub-volumes of clusters. This happens whenever short-lived ultra-relativistic
electrons are injected at a shock wave travelling through the ICM. The
electrons usually lose their energy before the shock wave can travel far
away. The electrons thereby produce synchrotron radio emission from a nearly
sheet-like volume. This emission likely forms the so-called giant radio relics
\citep{1998AA...332..395E, 1999ApJ...518..603R}. High resolution radio
polarisation maps of the relic in Abell 2256, which is one of the largest radio
relics known, reveal that the magnetic fields are organised in filaments or
sheets with an aspect ratio of at least 5 \citep{ClarkeEnsslin06}. Furthermore,
the Faraday rotation map of 3C465 reveals stripy patterns
\citep{2002ApJ...567..202E}, also suggesting the existence of intermittent
fields in galaxy clusters.  The presence of thermally isolated elongated cool
H-$\alpha$ filaments in the core of the Centaurus cluster is also best
understood by the existence of filamentary magnetic fields in that environment
\citep{2005MNRAS.363..216C}.

\subsection{Structure of the paper}

The paper is structured as follows: In Sect.~\ref{sec:sig} we introduce the
turbulent dynamo and summarise its expected and observed signatures in galaxy
clusters, mainly based on the abovementioned estimate of the magnetic
power-spectrum in the Hydra~A cluster core. In Sect.~\ref{sec:MTurb} we develop
a steady-state analytical description of the expected magnetic turbulence due
to the stirring motion of buoyant radio bubbles from the central galaxy, which
is assumed to regulate the energy content of the cool core. This is applied in
Sect.~\ref{sec:application} to a number of prominent cool core clusters, and
compared to existing information on magnetic fields whenever
available. Sect.~\ref{sec:concl} contains our conclusions.

Throughout the paper, we assume a Hubble constant of $H_0 = 70
\,\mathrm{km/s/Mpc}$, and translate literature values of luminosities, length
scales, electron densities and magnetic field strengths derived from Faraday
rotation measurements to this value.

\section {Turbulent magnetic dynamo\label{sec:sig}}
\subsection{Dynamo concepts}

The Kolmogorov-like magnetic power spectrum in the cool core of the Hydra~A
cluster indicates that the magnetic fields are shaped and probably amplified by
hydrodynamical turbulence \citep[e.g.][]{1992ApJ...386..464D}. Therefore, it
seems most promising to seek the origin of the observed magnetic power spectrum
in the theories of turbulent dynamos. A very similar view and approach are used
in the recent work by \citet{2006MNRAS.366.1437S}, in which the magnetic fields
in non-cooling core clusters were also assumed to be maintained by
turbulence. In that environment the turbulence is due to merger events and
galaxy motion, whereas here it is due to the inflation and buoyant motion of
radio bubbles.

Since cool cores of galaxy clusters are not believed to rotate, the gas flow is
probably non-helical and the galactic dynamo theories do not apply
here. Instead, the non-helical turbulent dynamo (also called small-scale
dynamo) should operate if the gas flow is sufficiently random, as it would be
in the case of developed turbulence. 

The physical details of the small-scale dynamo are still debated.
However, as it turns out, the exact nature of the small scale dynamo
is of minor importance for an understanding of cool core magnetic
fields. The observed magnetic field is probably determined by the
saturated state of such a dynamo, provided the dynamo had sufficient
time to amplify seed fields to the dynamically relevant strength. This
condition is fulfilled if either the dynamo is very efficient,
permitting the very weak primordial fields to be amplified, or if the
dynamo can start in an existing, relatively strong magnetisation,
which is only a few orders of magnitude below the saturation
level. There are arguments in favour of both pre-conditions being
fulfilled, which we discuss briefly.

The random gas motion would stretch and fold any initial seed magnetic fields
and lead to an exponential growth of the magnetic energy density with time with
the characteristic time-scale being the eddy-turnover time. This proceeds as
long as the dynamical back-reaction of the magnetic field is unimportant.  The
folding operations of the flow form small-scale magnetic reversals
perpendicular to the local field directions
\citep{2002PhRvE..65a6305S}.  Magnetic diffusivity limits the scales to be of the
order of $\lB \sim \lT \, R_\mathrm{m}^{-1/2}$, where $\lT$ is the turbulence
injection scale, and $R_\mathrm{m}$ is the magnetic Reynolds-number
\citep{1989MNRAS.241....1R}. The typical bending radius of the fields should be
of the order of the turbulence length-scale $\lT$.

This picture has been criticised by \citet{1993ApJ...411..518G} as being unable
to explain cluster magnetic fields. Their main objection is that the transverse
size of the magnetic structures $\lB$ should be extremely small, since the
magnetic Reynolds number is typically $R_\mathrm{m}\sim 10^{28\ldots 29}$ in a
cluster environment, leading to $\lB$ of the order of a light second. However,
magnetic structures with kpc size are required to accommodate the observed
Faraday rotation values with physically plausible cluster magnetic fields. This
reasoning lead \citet{1993ApJ...411..518G} to argue for a galactic origin of
the intra-cluster magnetic fields. Rough estimates of the magnetisation of the
intra-cluster medium indicate that galactic outflows in magnetised winds and
relativistic plasma jets should lead to a substantial seed
magnetisation\footnote{\citet{1987QJRAS..28..197R},
\citet{1990ApJ...364..451D}, \citet{1994Natur.368..434C},
\citet{1997ApJ...477..560E, 1998AA...333L..47E}, \citet{1999ApJ...511...56K},
\citet{2000ApJ...541...88V}, \citet{2001ApJ...560..178K}, \citet{bertone06}}.

The small-scale dynamo picture can only be reconciled with the observations if
the huge magnetic Reynolds number is replaced by a much lower effective
value. It was pointed out by \citet{1999PhRvL..83.2957S} that the gas motions
induced by the magnetic forces lead to a diminishing of the magnetic field
strength, which can be expressed approximately as an effective magnetic
diffusivity. This diffusivity increases with growing magnetic field strength,
leading to a decreasing effective magnetic Reynolds number permitting the
magnetic structures to grow to larger spatial scales. On larger scales more
turbulent energy density is available in a Kolmogorov cascade, thus larger
field strengths can be accommodated. Larger field strengths imply a further
decreased effective magnetic Reynolds number and therefore a further growth of
the fields in strength and length scale. This process continues until a
saturated state is reached. The magnetic $e$-folding time is of the order of
the turbulent eddy time scale and might be sufficient to amplify even
primordial magnetic fields in a cluster environment to their observed strength.

It was argued by \citet{2005ApJ...629..139S} that plasma instabilities
should lead to an accelerated regime of magnetic amplifications during
the pre-saturation phase. Assuming an effective description for the
plasma particle pitch angle scattering by the plasma-waves generated
in the instabilities \citet{2005ApJ...629..139S} showed that a nearly
explosive production of magnetic fields from very weak primordial seed
fields on cosmologically negligible times-scales might be possible.

Thus, although the details of the generation of dynamically relevant magnetic
fields in galaxy clusters are still unclear, it seems that there are sufficient
sources of magnetisation, and sufficiently efficient dynamo mechanisms present
to amplify such fields in cosmologically short times. The question for the
interpretation of the observational data is in which state do we expect the
fields to be at present. This requires an examination of the saturated dynamo
state.

\subsection{Saturated dynamo state}

We start our investigation of the saturated state with the view introduced by
\citet{1999PhRvL..83.2957S}.  As soon as the Lorentz force becomes sufficiently
strong, the fields do not follow the flow passively, but try to disentangle
themselves. This back-reaction motion leads to an increase in the effective
magnetic diffusivity, and therefore to a lower, renormalised magnetic Reynolds
number implying that fields become organised on larger scales. The Reynolds
number decreases until it reaches a critical value $\Rc \sim 10-100$, below
which the turbulent dynamo would stop operating. The system reaches a saturated
state, in which the magnetic correlation length scale
  \begin{equation}
\lB \sim \lT \, \Rc^{-1/2}    
  \end{equation}
is solely determined by the turbulence length scale and the critical Reynolds
number.

\citet{2006astro.ph..1246S} acknowledge that their explosive dynamo alone fails
to explain the observed level of cluster magnetisation, but only by up to one
order of magnitude. Therefore, the remaining amplification and the saturated
state should result from a conventional non-helical dynamo
\citep[e.g.][]{1999PhRvL..83.2957S}. However, \citet{2006astro.ph..1246S}
speculate that the physics of the plasma instability driven dynamo still
imprints on the saturated state, leading to a characteristic field strength
with $\lB \sim \lT \, (v_\mathrm{th, i}/v_\mathrm{T})^{1/2}\,
(\rho_\mathrm{i}/\lT)^{1/8}$. Here, $v_\mathrm{th}$ and $v_\mathrm{T}$ are the
ion and the turbulent velocity, respectively. $\rho_\mathrm{i}$ is the
gyroradius of an ion within a magnetic field that is in equipartition with the
turbulent energy density.  This picture of the saturated state can be
mathematically mapped onto the model of \citet{1999PhRvL..83.2957S} if
we define an effective magnetic Reynolds number of
\begin{equation}
\label{eq:Rc*}
 \Rc^* \sim \frac{v_\mathrm{T}}{v_\mathrm{th, i}}\, \left(
 \frac{\lT}{\rho_\mathrm{i}} \right)^\frac{1}{4}  
\end{equation}
which takes values in the range $10^{2\ldots 3}$ for cluster environments.

In the picture initially developed by \citet{Batchelor1950,
kazantsev67, 1990alch.book.....Z}, the magnetic fields are highly
intermittent: only a small fraction $f_B$ of the volume is actually
strongly magnetised, whereas the remaining volume $1-f_B$ does not
carry dynamically important magnetic
fields. \citet{1990alch.book.....Z} and others
\citep{1989MNRAS.241....1R, 1990IAUS..140..499S, 1999PhRvL..83.2957S}
assumed the magnetised regions to be organised as magnetic flux ropes
of volume $\lT\,\lB^2$ per eddy volume $\lT^3$. The occupied volume
fraction would be $f_B \sim {\lB^2}/{\lT^2} \sim \Rc^{-1}$. Such
highly filamentary structures were not confirmed by numerical
simulations. By using a spectral MHD code, \citet{2000ApJ...538..217C}
find for example that the magnetic autocorrelation functions exhibit
only a moderate anisotropy on scales where the magnetic energy density
peaks. However, numerical simulations in physical space indicate that
the magnetic structures are more sheet-like with thickness $\lB$
\citep{2004astro.ph..5052B, 2006astro.ph..1246S}, leading to a $f_B
\sim {\lB}/{\lT} \sim \Rc^{-1/2}$ as assumed in
\citet{2006MNRAS.366.1437S}. Furthermore, as we explained in
Sect.~\ref{sec:magobs}, radio observations of the cluster radio relic
in Abell 2256 also strongly support the assumption of intermittent
magnetic fields, probably of sheet-like structure.

In order to allow for any geometry of the magnetic structures in the saturated
state, we assume that they are effectively $d$-dimensional, giving
\begin{equation}
  f_B \sim \frac{\lT^d\,\lB^{3-d}}{\lT^3} \sim
  \Rc^{-\frac{3-d}{2}}\,,\;\mathrm{with}\; 1\le d\le 3.
\end{equation}

Under the steady-state conditions of the saturated dynamo, the
hydrodynamical dissipation of turbulent energy on scale $\lT$ and the relaxation
of magnetic structures bent on the same scales $\lT$ should have the same time
scales:
\begin{eqnarray}
\label{eq:tauStST}
  \tau_B \sim & \frac{\lT}{\mathrm{v}_\mathrm{A}^\mathrm{struct.}} & = 
  \lT \sqrt{\frac{\rho}{2\, \eps_{B}^\mathrm{struct.}}}\\
  \tau_\mathrm{T} \sim & \frac{\lT}{\vT} & = 
  \lT \sqrt{\frac{\rho}{2\, \epsT}} \,,
\end{eqnarray}
where $\mathrm{v}_\mathrm{A}^\mathrm{struct.}$ and $\eps_{B}^\mathrm{struct.}$ are the
Alfv\'enic velocity and energy density within the magnetic structures, and $\vT$
and $\epsT$ the turbulent velocity and energy density, respectively.
Then $\tau_B\sim \tau_\mathrm{T}$ implies that the magnetic fields within the
structures are in equipartition with the environmental turbulent energy
density, and the volume-averaged magnetic energy density is therefore lower
than these by the magnetic volume filling factor:
\begin{equation}
\label{eq:epsBfromEpsT}
  \eps_{B} \sim \eps_{B}^\mathrm{struct.}\,f_B \sim \epsT\,f_B \sim \epsT\,\Rc^{-\frac{3-d}{2}}\,.
\end{equation}
Thus, with the knowledge or assumption of the critical Reynolds number $\Rc$,
and the effective dimensionality  $d$ of the magnetised regions
it is possible to translate properties of the hydrodynamical turbulence to the
magnetic turbulence and vice versa, under the assumption that the system is in
the saturated dynamo state.

In the numerical examples, we will investigate three scenarios, which differ
  in assumed magnetic topology and critical magnetic Reynolds number: 
\begin{enumerate}
  \item Magnetic fields mostly organise in flux tubes ($d=1$, $\Rc = 20$,
  $\Rightarrow f_B = 0.05$). This scenario reflects the original assumption on
  turbulent magnetic structures by \citet{1990alch.book.....Z} but modified
  with the concept of the renormalised Reynolds number
  \citep{1999PhRvL..83.2957S}. Here we consider it to investigate the induced
  hydrodynamical viscosity on large scales, since there are published
  predictions for this quantity in the flux rope scenario
  \citep{2003ApJ...599..661L}.
  \item Magnetic fields mostly organise in sheets and ribbons ($d=2$, $\Rc =
  35$, $\Rightarrow f_B = 0.17$). This scenario is more in agreement with
  recent numerical simulations, as argued by
  \citet{2006MNRAS.366.1437S}.\footnote{\citet{2004PhRvE..70a6308H} numerically
  investigate magnetic turbulence in the case of a Prandtl number of order
  unity, and find a magnetic to turbulent energy ratio of the order of $f_B
  \approx 0.4$ in this regime. Since they report a $\Rc \approx 35$, their case
  could be described by $d = 2.5$ dimensional magnetic intermittency.  However,
  as \citet{2005ApJ...622..205D} point out, the dissipation rate of turbulent
  energy seems to be doubled compared to not (or less) magnetised scenarios,
  which would lead to a reduction in the turbulent and magnetic energy
  densities by a similar factor. For the observables of the magnetic fields,
  like field strength, length and Faraday signal, there would be very little
  numerical difference to the $d=2$ scenario. For this reason, we do not follow
  this scenario separately, but assume it to be subsumed under the $d=2$
  case.}
  \item The assumed saturated state of the explosive dynamo of
  \citet{2006astro.ph..1246S}, which we characterise by $d = 2$ dimensions
  and by an effective Reynolds number $\Rc = \Rc^*$ provided by Eq.~\ref{eq:Rc*}.
\end{enumerate}
The scaling of any quantity with the parameters $\Rc$ and $d$ will also be
provided, so that this theory may be applicable even if the adopted values need
readjustment.

\subsection{Magnetic viscosity}

In the saturated dynamo state, the magnetic fields do not passively
follow the hydrodynamical flow of the bulk motion, but possess an
independent velocity component due to the Lorentz force. This should
lead to a slippage of strongly and
weakly magnetised regions relative to each other, possibly causing
some friction due to induction of small scale eddies in the wakes of
the magnetic structures. For the flux rope dynamo scenario,
\citet{2003ApJ...599..661L} give the expected viscosity on large
spatial scales in the saturated state to be of the order of $4\%$ of
the turbulent diffusivity $\kappa_\mathrm{T} \approx \vT\, \,\lT/3$,
thus
\begin{equation}
\label{eq:visco}
\kappa_\mathrm{visc} \approx 0.04\, \frac{\vT\, \lT}{3}\,.
\end{equation}
Also in the scenario of mainly 2-dimensional magnetic structures, one would
expect such viscous effects, but no estimates of the viscosity exist to our
knowledge.

\subsection{Signatures}

According to the scenario described above, the magnetic fields in the turbulent
cool cores of galaxy clusters should exhibit the following properties:
\begin{itemize}
\item[A.] The magnetic power spectrum should either reflect the
  hydrodynamical power spectrum due to their dynamical coupling, just with
  a lower normalisation and on smaller spatial scales or the small-scale
  magnetic fluctuations are expected to follow a Goldreich-Sridar law, which is
  indistinguishable from a Kolmogorov spectrum in the isotropic average
  \citep{1997ApJ...485..680G}. That means that a Kolmogorov-inertial range
  power law behaviour of hydrodynamical turbulence should also be found in the
  magnetic spectrum in any scenario.
\item[B.] The average magnetic energy density $\varepsilon_B$ is lower than the
turbulent kinetic energy density $\varepsilon_{\rm kin}$ by $\varepsilon_B
\approx \varepsilon_{\rm kin} \, f_B$, where $f_B \sim \Rc^{-(3-d)/2} \sim 0.05
\ldots 0.2$ depending on the critical magnetic Reynolds number and the
dimensionality of the magnetic structures.
\item[C.] The magnetic fluctuations are concentrated on a perpendicular scale
$\lB$, which is smaller than the hydrodynamical turbulence injection scale
$\lT$ by $\lB \approx \lT \Rc^{-1/2}$.
\item[D.] Magnetic correlations exist up to a scale $\lT$, turn there into an
  anti-correlation as a consequence of $\vec{\nabla}\cdot\vec{B} = 0$, and
  quickly decay on larger scales.
\item[E.] The fields may be spatially intermittent. This may be understood by
  Zeldovich's flux rope model, in which magnetic ropes with diameter $\lB$ are
  bent on scales of the order $\lT$. Alternatively, as also supported by recent
  numerical simulations, it can be understood in terms of magnetic sheets of
  thickness $\lB$ and size $\lT$.
\item[F.] In the case of intermittent fields, the field strength within the
  magnetic structures should be in energy equipartition with the average
  turbulent kinetic energy density of their environment.
\item[G.] The magnetic drag of such intermittent structures produces a
  hydrodynamical viscosity on large scales. In case of the Zeldovich-flux rope
  scenario, an estimate of a viscosity of $4\%$ of the turbulent diffusivity
  $\kappa_\mathrm{T} \approx \vT\, \,\lT/3$ was made by
  \citet{2003ApJ...599..661L}.
\end{itemize}

\subsection {Observations}

We investigate briefly if the above predictions of the non-helical dynamo
theory are in agreement with observations.
\begin{itemize}
\item[A.] A Kolmogorov-like magnetic power spectrum in a cluster cool core is
  revealed by the Faraday rotation map of Hydra~A \citep{2005A&A...434...67V}.
\item[B.] Translating the Faraday rotation based estimate of the magnetic energy
  density to the expected turbulent energy density $\epsT \sim \eps_B/f_B$ in
  the Hydra~A cluster yields $0.3\ldots 1.0 \,\cdot\, 10^{-10}\, {\rm erg\, cm^{-3}}$,
  which corresponds to turbulent velocities of $v_{\rm turb} \approx
  250\ldots 430 \,{\rm km/s}$. This is comparable to
  velocities of buoyant radio plasma bubbles \citep{2002A&A...384L..27E}, which
  are expected to stir up turbulence \citep[e.~g.~][]{2001ApJ...554..261C}.
\item[C.] The expected turbulence injection scale in the Hydra~A cluster core
  is of the order of $\lT\sim \lB\,\Rc^{1/2} \sim 10 \ldots 20$ kpc, again consistent
  with the radio plasma of Hydra~A being the source of turbulence,
  since the turbulence injection scale and the radio lobes of Hydra A
  have comparable dimensions. The dynamical connection of the radio source
  length scale and the magnetic turbulence scale would explain why the Faraday
  map of Hydra~A is conveniently sized to show us the peak of the magnetic
  power spectrum (see Fig.~\ref{fig:BturbHydra}).
\item[D.] The expectation of weak magnetic fluctuations on scales larger than
  the turbulence injection scale is hard to test with the available data due to
  the limited size of the \RM map used. However, the downturn of the magnetic
  power spectrum at small $k$-values visible in Fig.~\ref{fig:BturbHydra} is in
  good agreement with the requirement of magnetic anti-correlations on large scales.
\item[E.] Magnetic structures in the form of flux ropes or ribbons might have been
  detected as striped patterns in the \RM map of 3C465
  \citep{2002ApJ...567..202E}, and as polarised synchrotron filaments in the
  cluster radio relic in Abell 2256 \citep{ClarkeEnsslin06}, in support of the
  assumed intermittence of the small-scale dynamo .
\item[F.] The fraction of the strongly magnetised volume may be as small as
  $f_{\rm B} = R_{\rm c}^{-1} \approx 0.05\ldots 0.2$. Strongly intermittent
  fields in galaxy clusters may help to reconcile the discrepancy between
  Faraday and inverse Compton based magnetic field estimates
  \citep{1999AA...344..409E}. The latter estimate could easily be biased to low
  field regions due to the faster removal of ultra-relativistic electrons in
  strong field regions by synchrotron emission.
\item[G.] The expected hydrodynamical viscosity on large scales in the Hydra
  cluster is of the order of $10^{28}\,{\rm cm^2/s}$ (if the estimate in
  Eq.~\ref{eq:visco} based on the flux rope picture is
  applicable). \citet{2003MNRAS.344L..48F} argue, for the comparable Perseus
  cluster cool core, for a lower limit on the large-scale viscosity of $4\cdot
  10^{27}\, {\rm cm^2/s}$. They base their arguments on the observation of very
  elongated H-$\alpha$ filaments which suggest a laminar flow pattern behind
  buoyantly rising radio bubbles. An upper limit on the viscosity in the
  (non-cooling core) Coma cluster of $\sim 3\cdot \,10^{29}\, {\rm cm^2/s}$ is
  reported by \citet{2004A&A...426..387S}. Both limits are consistent with our
  coarse estimate of the large scale viscosity and enclose it. However, the
  presence of a significant viscosity, as well as the underlying magnetic
  intermittency, in galaxy clusters should be regarded as being speculative,
  and not directly confirmed by observations. Nevertheless, there is a growing
  number of theoretical investigations\footnote{\citet{1989MNRAS.239..479P,
  2001ApJ...554..261C, 2002MNRAS.332..271R, 2004ApJ...600..650F,
  2004ApJ...615..675R, 2005ApJ...622..205D, 2005MNRAS.357..242R,
  2005ApJ...630..740B, 2005MNRAS.359..493K}} on the possibility that viscosity
  helps to explain properties of the cluster gas.
\end{itemize}

\section{Magneto-hydrodynamic turbulence in cool cores\label{sec:MTurb}}
\subsection{Scope of the approach}

A theoretical model of the magneto hydrodynamic turbulence in cool
cores of galaxy clusters is developed in the following. The model is
intended to capture the most essential features of the physical
picture. We do not attempt to make accurate numerical predictions, but
hope to get insight into the scaling relations of the different
quantities and their order of magnitude values.

Although cool cores exhibit a structured gas distribution, for our
order-of-magnitude calculation we describe them as quasi-homogeneous spheres.
We briefly present a radial extension of our model in
Sect.~\ref{sec:RMprofile}.

The episodic injection of radio plasma by the central galaxy
into cool cores should lead to temporal variations of the state of the
core. Nevertheless, we treat the system as being in a steady
state. Our estimates provide therefore only approximate temporal mean
values. Observations of cool cores will always show snapshots of the
cool core evolution. Thus, deviations between our predictions and the
observed state of individual cool cores should not be too surprising. 

Detailed numerical modelling of the processes would be required to overcome our
simplifying assumptions which is well beyond the scope of this initial
investigation of the scenario presented.

\subsection{A simplified cool core description}

A cool core is a condensation of cold gas of mass $\Mcc$, which
dropped out of the hot phase of a galaxy cluster due to the faster
radiative cooling of denser gas. The observables of the
cool core, which can be used as diagnostics of the physical
parameters, are the bolometric X-ray luminosity of the cool core
$\Lcc$, its temperature $\Tcc$, and its radius $\Rcc$ from which the
central electron density $\ncc$ can be deduced.

The energy feedback of the central radio source prevents the cool core gas from
falling below a characteristic temperature\footnote{We express temperatures in
terms of energies by setting the Boltzman constant to unity ($k_\mathrm{B}
=1$)} $\Tcc \sim \,$keV, which is given by the requirement that the atomic line
emission does not dominate. Otherwise, in the case of a lower temperature, the
gas in the cool core would rapidly condense onto the central galaxy due to a
cooling catastrophe driven by emission-lines.  The pressure in a cool core is
\begin{equation}
   \Pcc = 2\, \ncc\, \Tcc\,,
\end{equation}
where we ignore here and in the following the presence of any element heavier
than hydrogen in pressure and mass terms for simplicity. The cool core gas
mass is therefore
\begin{equation}
  \Mcc = \ncc \, \mp \, \Vcc \,,\; \mbox{where}\;  \Vcc =  \frac{4\, \pi}{3} \Rcc^3\, .
\end{equation}
The X-ray luminosity of the cool core is given by
\begin{equation}
\label{eq:Lcc}
   \Lcc = \LX(\Tcc) \, \ncc^2 \, \Vcc \,,
\end{equation}
where 
\begin{equation}
   \LX(\Tcc) = \L0\, \Tcc^\frac12 + \Lambda_\mathrm{lines}(\Tcc)
\end{equation}
consists of a Bremsstrahlung- and a line emission term. The line emission term
should be at best comparable to the Bremsstrahlung-term, since otherwise the
full core would run into a cooling catastrophe. Therefore, we can ignore the
line-term in our rough estimate
\begin{equation}
   \LX(\Tcc) \approx \L0\, \Tcc^\frac12 \,,
\end{equation}
with $\L0 = 5.96\cdot 10^{-24}\,\mbox{erg}\, \mbox{s}^{-1}\,
\mbox{cm}^{3}\, \mbox{keV}^{-1/2}$ (assuming a metalicity of 0.3 solar). 

\subsection{Hydrodynamical turbulence injection\label{sec:hti}}

We assume that the injection from radio galaxies is the dominant heating
mechanism.  Although radio galaxies inject energy, not all of it is in the form
of turbulence. During the initial phase of inflation of a radio bubble there
are shocks, which can heat the environment \citep{1998ApJ...501..126H,
2003MNRAS.344L..43F}. Later, a series of sound waves are produced, which might
be dissipated through viscosity in the
medium\footnote{\citet{1989MNRAS.239..479P, 2001ApJ...554..261C,
2002MNRAS.332..271R, 2004ApJ...600..650F, 2004ApJ...615..675R,
2005ApJ...622..205D}}. When a radio bubble raises buoyantly in the cool core
atmosphere, the environmental gas flows around it, which leads to
injection of kinetic energy from the central radio galaxy into the cool
core. The time averaged turbulence power of the source is
\begin{equation}
  L_\mathrm{T} = \fT\, \Lcc = \etaT \, L_\mathrm{rg},
\end{equation}
where $\fT$ is the fraction of cool core heating by turbulence, $\etaT\le 1$ is
the efficiency of kinetic energy transfer between the injected radio plasma and
the cool core gas, and $L_\mathrm{rg}$ is the mechanical luminosity of the
radio galaxy ($P\,dV$ work per time). \citet{2002MNRAS.332..729C} argue that
the efficiency factor $\etaT$ should be of the order of one since the radio
plasma loses most of its energy by adiabatic expansion during its buoyant rise
through one scale-height of the cluster atmosphere. This energy is transfered
mechanically to the kinetic energy of the ICM gas and is finally dissipated as
heat. It is controversial whether most of this energy is dissipated within the
cool core, or within the outer regions of the cluster. Here, we assume the
former and set $\etaT = \fT= 0.5$ also allowing non-turbulent heating (shock
waves, sound waves, heat transport). However, we also provide the scaling of
our results with these parameters.

The injected kinetic energy is dissipated through a Kolmogorov cascade within
an eddy turnover-time
\begin{equation}
  \tau_\mathrm{T} \sim \lT/\vT\,,
\end{equation}
where $\lT$ and $\vT$ are the turbulence injection
scale and turbulence root mean square velocity, respectively. The average
turbulent energy of the cool core $E_\mathrm{T} = \Mcc\,\vT^2/2$ is
therefore
\begin{equation}
  E_\mathrm{T} = L_\mathrm{T}\,\tau_\mathrm{T} \approx {\fT\,\Lcc
  \,\lT}/{\vT}\,,
\end{equation}
yielding
\begin{equation}
\label{eq:turbBalance}
  \frac{\vT^3}{\lT} = \frac{2\,\fT\,\Lcc}{\Mcc} =
  \frac{2\,\fT\,\Lcc}{\mp \,\ncc\, \Vcc}\,.  
\end{equation}
This would allow the determination of the turbulent velocity and energy
density if the turbulent injection scale were known. In the
following, we attempt a rough estimate of the expected turbulent length
scale $\lT$.

\subsection{Hydrodynamical turbulence injection scale}

In the picture adopted for this work, the turbulence is stirred by the
movement of buoyant radio plasma bubbles. The turbulence injection scale
should therefore be of the order of the radius $r_\mathrm{bub}$ of the
bubbles, which we approximate to be spheres. Assuming an
ultra-relativistic equation of state for the radio plasma, we find the
volume of the bubble as
\begin{equation}
\label{eq:Vbub}
 V_\mathrm{bub} = E_\mathrm{bub}/(4\,P)\,,
\end{equation}
where 
\begin{equation}
\label{eq:Ebub}
  E_\mathrm{bub} = L_\mathrm{rg}\, \tau_\mathrm{bub}/2 = \fT
  \Lcc \, \tau_\mathrm{bub}/(2\,\etaT)
\end{equation}
is the mechanical energy released by the radio galaxy into a bubble during the
time $\tau_\mathrm{bub}$ the bubble needs to leave the cool core buoyantly. If
the turbulence stirring bubbles were filled mostly by thermal gas, the volume
of the bubble would be $V_\mathrm{bub} = 2\,E_\mathrm{bub}/(5\,P)$, which would
make some difference to our estimates, but none that change the order of
magnitude of our results.

The time available for the jets to inflate the bubble is its rise time through
the cool core
\begin{equation}
\label{eq:tbub}
  \tau_\mathrm{bub} \sim \frac{\Rcc}{v_\mathrm{bub}} \sim 
  \left( \frac{\Rcc}{r_\mathrm{bub}} \right)^{\frac{1}{2}} \,
  \frac{\Rcc}{2\, c_\mathrm{s}}\,,
\end{equation}
and can be estimated from the balance of drag and buoyance forces
\citep[e.g.][]{2002A&A...384L..27E}.

Combining Eq.~\ref{eq:Vbub},~\ref{eq:Ebub}, and~\ref{eq:tbub} yields
the bubble radius 
\begin{equation}
\label{eq:rbub}
  r_\mathrm{bub} = \left( \frac{3\, \fT\, \Lcc\,
  \Rcc^{3/2}}{64\, \pi\, \etaT\, c_\mathrm{s}\, \Pcc} \right)^{\frac{2}{7}}\,,
\end{equation}
and thereby the required hydrodynamical turbulence scale
$\lT \sim r_\mathrm{bub}$.  Using this and the derived relations above,
the turbulent velocity is
\begin{eqnarray}
  \vT &=& \frac{3^{\frac{1}{21}}   \,
  \etaT^{\frac{1}{3}} \, \L0^{\frac{3}{7}} \,
  \ncc^{\frac{3}{7}} \, \Rcc^{\frac{3}{7}} \,
  \Tcc^{\frac{1}{14}}}{2^{\frac{4}{21}}   \, 5^{\frac{1}{21}} \,
  \fT^{\frac{1}{3}} \, \mp^{\frac{2}{7}} } \\ 
&=& 173\, \mathrm{km/s}\,  
\left(\frac{\etaT}{\fT}  \right)^{\frac{1}{3}} 
\left(\frac{\ncc}{\mathrm{0.1\, cm^{-3}}}  \right)^{\frac{3}{7}} 
\left(\frac{\Rcc}{\mathrm{10\, kpc}}  \right)^{\frac{3}{7}}
\left(\frac{\Tcc}{\mathrm{keV}}  \right)^{\frac{1}{14}}\,,
\nonumber
\end{eqnarray}
%
and thus, the turbulent energy density is given by
\begin{eqnarray}
  \epsT &=& \frac{ 3^{\frac{2}{21}}  \,
  \etaT^{\frac{2}{3}} \, \L0^{\frac{6}{7}}\, \mp^{\frac{3}{7}} \,
  \ncc^{\frac{13}{7}} \, \Rcc^{\frac{6}{7}} \,
  \Tcc^{\frac{1}{7}}}{2^{\frac{29}{21}}   \,5^{\frac{2}{21}} \, \fT^{\frac{2}{3}} } \,.
\end{eqnarray}

Table~\ref{tab:cc} contains the values for $\vT$, $\lT = r_\mathrm{bub}$, and
$\epsT$ as for a number of prominent cool core clusters, expected from our
steady-state description of cool core turbulence. Deviations due to episodic
evolution of cool core turbulence are possible and expected.

\subsection{Magnetic turbulence}

The energy density of the magnetic turbulence $\eps_B$ is lower by a factor
$f_B \sim 0.05\ldots 0.2$ than the kinematic one $\epsT$. The length-scale is
also smaller by a factor $\Rc^{-1/2}$. The root-mean-square magnetic field
strength in the cool core is therefore given by
\begin{eqnarray}
  B_\mathrm{rms} &=& \sqrt{8\,\pi\, \epsT\, f_B} =  
\frac{2^{\frac{17}{21}}   \, 3^{\frac{1}{21}} \pi^{\frac{1}{2}}  \,
  \etaT^{\frac{1}{3}} \, \L0^{\frac{3}{7}}\, \mp^{\frac{3}{14}} \,
  \ncc^{\frac{13}{14}} \, \Rcc^{\frac{3}{7}} \,
  \Tcc^{\frac{1}{14}}}{5^{\frac{1}{21}} \, \fT^{\frac{1}{3}} \,
  \Rc^{\frac{3-d}{4}}  } \\
&=& \left\{
\begin{array}{rl}
\!\!
5.6\, \mu\mathrm{G} &  
\left(\frac{\etaT}{\fT}  \right)^{\frac{1}{3}} 
\left(\frac{\ncc}{\mathrm{0.1\, cm^{-3}}}  \right)^{\frac{13}{14}} 
\left(\frac{\Rcc}{\mathrm{10\, kpc}}  \right)^{\frac{3}{7}}
\left(\frac{\Tcc}{\mathrm{keV}}  \right)^{\frac{1}{14}}
\left(\frac{\Rc}{20} \right)^{-\frac{1}{2}}\\
\!\!
10\, \mu\mathrm{G} & 
\left(\frac{\etaT}{\fT}  \right)^{\frac{1}{3}} 
\left(\frac{\ncc}{\mathrm{0.1\, cm^{-3}}}  \right)^{\frac{13}{14}} 
\left(\frac{\Rcc}{\mathrm{10\, kpc}}  \right)^{\frac{3}{7}}
\left(\frac{\Tcc}{\mathrm{keV}}  \right)^{\frac{1}{14}}
\left(\frac{\Rc}{35} \right)^{-\frac{1}{4}}
\end{array}
\right.
\nonumber
\end{eqnarray}
in our flux rope ($d=1$) and magnetic sheet ($d=2$) scenarios, respectively. The magnetic autocorrelation length
is
\begin{eqnarray}
  \lambda_B &=& \frac{\lT}{\Rc^{\frac{1}{2}}} = \frac{3^{\frac{1}{7}}
  \, \L0^{\frac{2}{7}}\, \mp^{\frac{1}{7}} \, \ncc^{\frac{2}{7}} \,
  \Rcc^{\frac{9}{7}}}{ 2^{\frac{11}{7}} \,5^{\frac{1}{7}} \,
  \Rc^{\frac{1}{2}} \, \Tcc^{\frac{2}{7}}} \\ &=& 0.53\,
  \mathrm{kpc}\, \left(\frac{\ncc}{\mathrm{0.1\, cm^{-3}}}
  \right)^{\frac{2}{7}} \left(\frac{\Rcc}{\mathrm{10\, kpc}}
  \right)^{\frac{9}{7}} \left(\frac{\Tcc}{\mathrm{keV}}
  \right)^{-\frac{2}{7}} \left(\frac{\Rc}{20} \right)^{-\frac{1}{2}}
  \nonumber
\label{eq:lmabdaB}
\end{eqnarray}

\subsection{Predicting an observable: RM dispersion}

Knowing the depth of the Faraday screen $\Rcc$, the magnetic field strength
$\langle B^2 \rangle$ and the magnetic autocorrelation length $\lambda_B$
allows one to predict the expected dispersion of the Faraday rotation measure
from a central radio source\footnote{Background sources should have twice the
variance $\langle RM^2 \rangle$ given here due to the doubled screen depth.},
assuming an isotropic distribution of magnetic field strengths with the help of
Eq. 40 of \citet{2003A&A...401..835E}:
\begin{eqnarray}
\label{eq:RMformula}
\langle RM^2   \rangle &=& \frac{1}{2}\, a_0^2\, \ncc^2\,
\Rcc\, \lambda_B \, \langle B^2   \rangle \\
\label{eq:RMformulaB}
 &=& \frac{3^{\frac{5}{21}} \pi \,a_0^2\,   \etaT^{\frac{2}{3}} \,
  \L0^{\frac{8}{7}}\, \mp^{\frac{4}{7}} \, 
  \ncc^{\frac{29}{7}} \, \Rcc^{\frac{22}{7}} }{
2^{\frac{20}{21}}   \,5^{\frac{5}{21}} \, \fT^{\frac{2}{3}} \,
  \Rc^{2-\frac{d}{2}} \,  \Tcc^{\frac{1}{7}} }   \\
RM_\mathrm{rms}^\mathrm{exp} &=&
\left\{
\begin{array}{rl}
\!\!
\frac{739}{\mathrm{m^2}} &
\left(\frac{\etaT}{\fT}  \right)^{\frac{1}{3}} 
\left(\frac{\ncc}{\mathrm{0.1\, cm^{-3}}}  \right)^{\frac{29}{14}} 
\left(\frac{\Rcc}{\mathrm{10\, kpc}}  \right)^{\frac{11}{7}}
\left(\frac{\Tcc}{\mathrm{keV}}  \right)^{-\frac{1}{14}}
\left(\frac{\Rc}{20} \right)^{-\frac{3}{4}}\\
\!\!
\frac{1180}{\mathrm{m^2}} &
\left(\frac{\etaT}{\fT}  \right)^{\frac{1}{3}} 
\left(\frac{\ncc}{\mathrm{0.1\, cm^{-3}}}  \right)^{\frac{29}{14}} 
\left(\frac{\Rcc}{\mathrm{10\, kpc}}  \right)^{\frac{11}{7}}
\left(\frac{\Tcc}{\mathrm{keV}}  \right)^{-\frac{1}{14}}
\left(\frac{\Rc}{35} \right)^{-\frac{1}{2}}\\
\end{array}
\right.
\nonumber
\end{eqnarray}
The two cases correspond to filamentary and sheet-like magnetic structure
scenarios, respectively.  Here, $a_0 = {e^3}/({2\pi \,m_e^2\,c^4})$ is the usual
Faraday rotation constant%
\footnote{Eq.~\ref{eq:RMformula} differs from the usually
used, but inaccurate \RM dispersion formula, which is based on the cell model
for the magnetic fields
\begin{displaymath}
\langle RM^2   \rangle \approx \frac{1}{3}\, a_0^2\, \ncc^2\,
\Rcc\, \lambda_{RM} \, \langle B^2   \rangle \;\;.
\end{displaymath}
The proper magnetic autocorrelation length $\lambda_B$ is used in
Eq.~\ref{eq:RMformula}, which differs in a non-trivial way -- since this is
magnetic power-spectrum dependent -- from the \RM autocorrelation length
$\lambda_\mathrm{RM}$ derived from \RM maps. Usually, $\lambda_{RM}> \lB$.
Second, the numerical factor $\frac{1}{2}$ in Eq.~\ref{eq:RMformula} properly
takes account of the effect of the constraint $\vec{\nabla}\cdot\vec{B} = 0$,
whereas for the above Eq., uncorrelated patches with internally constant
magnetic fields were assumed, which do not have $\vec{\nabla}\cdot\vec{B} = 0$
at the patch boundaries.  Due to the differences, published magnetic field
values based on the above Eq.  do not need be accurate, although the two errors
partly compensate each other.}.

The scaling of the Faraday dispersion, which can be written approximately as
${RM}_\mathrm{rms}^\mathrm{exp} \propto \ncc^2\, \Rcc^{3/2} \, \Tcc^0$, is
sufficiently similar to that of the bolometric luminosity of the cool core
$\Lcc \propto \ncc^2\, \Rcc^3 \, \Tcc^{1/2}$ and also to the mass deposition
rate $\dot{M} \propto \Lcc/\Tcc \propto \ncc^2\, \Rcc^3 \, \Tcc^{-1/2}$ to
expect significant correlations, which are indeed observed
\citep[e.g.][]{2002MNRAS.334..769T}.

\begin{figure}[t]
\begin{center}
\resizebox{\hsize}{!}{\includegraphics{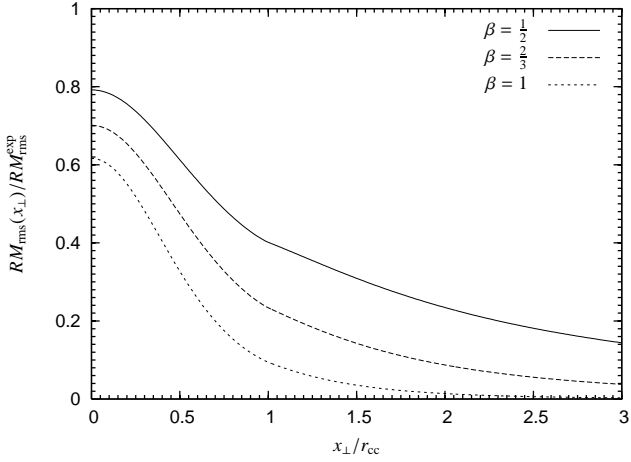}}
\end{center}
\vspace{-0.0cm}
\caption[]{Radial profile of the expected \RM dispersion for a line of sight
  starting in the midplane of a $\beta$-profile cool core, but displaced from
  the cluster centre by $x_\perp$. The \RM dispersion is given in units of our
  result $RM_\mathrm{rms}^\mathrm{exp}$ in Eq.~\ref{eq:RMformula} for the solid
  sphere model of the cool core.}
\label{fig:RMprofile}
\end{figure}

\subsection{Approximate treatment of the cool core structure\label{sec:RMprofile}}

In the following, we generalise the simplified geometry of a homogeneous cool
core to one which has a radial structure. This generalisation allows us to predict
the radial dependence of the various quantities and to test if our predictions
of the central \RM dispersion are expected to change significantly due
to the contribution from radii outside the cool core radius.

In order to provide a radius dependent geometry, we make an additional
assumption about the radial balance between heating and cooling. We
assume that the cool core is settled in an hydrodynamical state,
where heating and cooling are balanced locally at all radii below the
cooling radius. Note that this does not necessarily need to be true,
since in a convecting system the location of heat deposition and
radiative cooling can differ, but for the sake of simplicity, we use
this approximation.

The cool core is divided into spherical shells of thickness $dr$, which have a
volume $dV(r) = 4\,\pi\,r^2\,dr$ and a luminosity $dL(r) = \LX(\Tcc) \, \ncc^2
\, dV(r)$. Our arguments about the balance of turbulent heating and
radiative cooling made in Sect.~\ref{sec:hti} can be identically
applied to
each of these shells individually, leading to a relation equivalent to
Eq.~\ref{eq:turbBalance}:
\begin{equation}
\label{eq:turbBalance(r)}
  \frac{\vT^3(r)}{\lT(r)} =  \frac{2\,\fT}{\mp \,\ncc}\,\frac{dL}{dV}(r)\,.  
\end{equation}

To derive the local turbulence injection scale, we use again the local
characteristic buoyancy time $\tau_\mathrm{bub}(r) \sim r/v_\mathrm{bub}(r)$ to
predict the bubble size at any cluster radius $r>\Rcc$. For $r\le \Rcc$, we use
$\Rcc$ to treat the central flat density profile properly.  In doing so, we
find that the replacement
\begin{equation}
  \Rcc \rightarrow \max(\Rcc, r)
\end{equation}
in any former equation describing local properties as in
Eqs.~\ref{eq:rbub}-\ref{eq:lmabdaB} gives the appropriate function of radius
$r$. Of special interest may be the expected radial scaling of the typical
magnetic field strength: $B_\mathrm{rms} \propto n_{\mathrm{e}}^{13/14}\!(r) \,
r^{3/7}$\, which is usually a decreasing function of the radius due to the
steepness of the electron profiles. For example for $n_{\mathrm{e}} \propto
r^{-2}$ (and $T(r) \approx const$) we find $B_\mathrm{rms} \propto r^{-10/7}
\propto n_{\mathrm{e}}^{5/7}$ which lies well within the range of usually
assumed magnetic field scaling with electron density $B_\mathrm{rms} \propto
n_{\mathrm{e}}^{0.5\ldots 1}$ \citep[e.g.][]{2001A&A...378..777D}.

Global quantities like the energy content, the luminosity of a
volume, etc. are calculated by performing the appropriate volume
integrals. The increase of the Faraday dispersion along the line of
sight should therefore be given by the differential analogy to
Eq.~\ref{eq:RMformula}

\begin{eqnarray}
  \label{eq:RMformula2}
\frac{d\langle {RM}^2   \rangle}{dl} &=& 
 \frac{1}{2}\, a_0^2\, \ncc^2\,
 \lambda_B \, \langle B^2   \rangle \\
  \label{eq:RMformula2B}
&=&\frac{3^{\frac{5}{21}} \pi \,a_0^2\,   \etaT^{\frac{2}{3}} \,
  \L0^{\frac{8}{7}}\, \mp^{\frac{4}{7}} \, 
  n_\mathrm{e}^{\frac{29}{7}}\!(r) \, \max^{\frac{15}{7}}\!(\Rcc,r)}{
2^{\frac{20}{21}}   \,5^{\frac{5}{21}} \, \fT^{\frac{2}{3}} \,
  \Rc^{2-\frac{d}{2}} \,  T_\mathrm{e}^{\frac{1}{7}}\!(r) }\,, 
\end{eqnarray}
which reduces to the former result given by Eq.~\ref{eq:RMformulaB} if
integrated over a cool core sphere with constant properties within
$\Rcc$.

Eq.~\ref{eq:RMformula2B} allows us to check under which conditions our constant
core model gives appropriate results, and under which conditions the \RM
contributions from larger radii become essential.

For an electron density profile described by the usual $\beta$-model,
\begin{equation}
  n_\mathrm{e}(r) = \ncc \left(1+\frac{r^2}{\Rcc^2}\right)^{-\frac{3}{2}\,\beta}\,,
\end{equation}
we find that the contribution to the \RM dispersion per logarithmic
radius peaks around the core radius $\Rcc$ if $\beta > 0.25$ (a similar
statement can be made for the cool core dominating the X-ray emissivity
if $\beta > 0.5$). Since typical cluster electron density profiles are
described by $\beta \ge 0.5$, our ignorance of contributions of outer regions
to the \RM dispersion only introduces a moderate error of the order of
$20-40$\% (depending on $\beta$) for lines of sight starting in the cluster
centre. Off-centre lines of sight, which often occur for extended radio lobes in
our sample, will result in significantly lower \RM dispersions. This can be seen
in Fig.~\ref{fig:RMprofile}, where numerically estimated \textit{RM}-profiles are
shown. 

Typical pairs of radio lobes are not located in the midplane of the
cluster. One lobe has a shorter and one has a longer line-of-sight through the
Faraday-active medium. In a spherical geometry of the cluster gas, the
additionally accumulated $\langle RM^2 \rangle$ of the back lobe, and the
missing of the front lobe should compensate roughly in an average over both
lobes, provided both lobes are in mirror symmetric positions with respect to
the cluster centre. Therefore, in such a case the midplane \RM dispersion is a
good reference point.

\section{Application to cool cores\label{sec:application}}

\begin{figure}[t]
\begin{center}
\resizebox{\hsize}{!}{\includegraphics{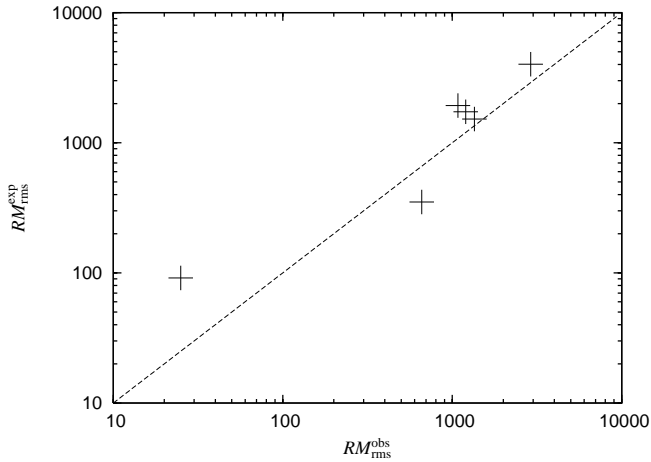}}
\end{center}
\vspace{-0.0cm}
\caption[]{Comparison of observed (cross) and theoretically expected (case
  $d=1$, $\Rc =20$) (dashed line) root-mean-squared \RM signal. The expected
  \RM is calculated for a radio source located at the centre of the cluster,
  whereas most \RM measurements are somewhat off-centre, probably leading to a
  reduced observed signal.}
\label{fig:RMRM} 
\end{figure}

In the following, we apply our model to a number of cool core
clusters. The input parameters of our calculations are the central
electron density $\ncc$, the central gas temperature $\Tcc$, and the
core radius $\Rcc$ of the cool core. The numbers are taken from the
literature and the corresponding references are given in the sections
refering to individual clusters in the Appendix~\ref{app:A}, together
with detailed discussions of the individual datasets.  The cluster
parameters and the derived properties are summarised in
Table~\ref{tab:cc}.

As a consistency check for the radii of the cool cores used, we calculate the
emissivity $\Lcc$ within the cool core according to Eq.~\ref{eq:Lcc} and
compare this to the reported bolometric X-ray luminosity of the complete
cluster $\mathrm{L}_{\mathrm{cluster}}$. The fraction of X-ray luminosity due
to the cool core is in the range of $\Lcc/\mathrm{L}_{\mathrm{cluster}} =
1\%-50\%$. The typical fractional luminosity within the cooling radius, which
is defined by the gas having a cooling time less than the Hubble time, was
estimated by \citet{1998MNRAS.298..416P}. Since the cooling radius is larger
than the core radius of the cool core, a systematic difference is expected, and
indeed found since $L(<r_\mathrm{cooling})/\mathrm{L}_{\mathrm{cluster}} =
40\%-60\%$ according to \citet{1998MNRAS.298..416P}.  The large differences for
some clusters between the ratios are due to shallow electron density profiles
with $\beta \approx 0.5$ (see Sect.~\ref{sec:RMprofile}).

The other quantities in Table~\ref{tab:cc} are calculated according to the
formulae given in Sect.~\ref{sec:MTurb}.  Note that our expected
${RM}_\mathrm{rms}^\mathrm{exp}$ is calculated for a polarised synchrotron
source in the middle of the cool core, whereas the real radio emitting volume
may be displaced due to an inclination between radio jet and line-of-sight,
and/or due to a non-central position. This will cause some deviation of our
expectations from the observations and usually biases the observational values
to be lower.

\begin{figure}[t]
\begin{center}
\resizebox{\hsize}{!}{\includegraphics{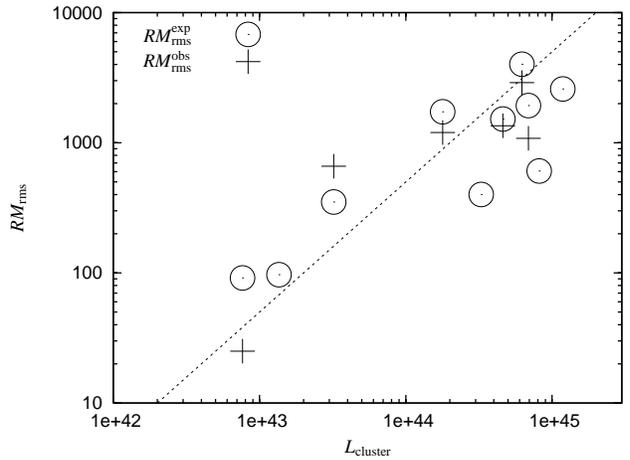}}
\end{center}
\vspace{-0.0cm}
\caption[]{Expected (case $d=1$, $\Rc =20$) and observed \RM
  dispersion values versus the bolometric cluster luminosity. A linear
  correlation is also shown to guide the eye.}
\label{fig:LXRM} 
\end{figure}

For a sample of cool cores with luminosities between
$10^{43}-10^{45}\,\mathrm{erg/s}$, we predict turbulent velocities in the range
$100-300 \,\mathrm{km/s}$, magnetic field strengths in the range $3-13 \,\mu$G
($1-d$ scenario) or $6-23 \,\mu$G ($2-d$ scenario), and Faraday dispersions in
the range $100-4000\, \mathrm{m^{-2}}$ ($150-6000$ in the $2-d$
scenario). These values should be compared to existing Faraday rotation
measurements, field strength estimates, and future X-ray-spectroscopically
determined velocity dispersions, as done in Table~\ref{tab:cc}, and
Figs.~\ref{fig:RMRM}~\&~\ref{fig:LXRM}. We further list our expectations for
the large-scale magnetic viscosity, which is speculative, and which were
estimated using the calculations of \citet{2003ApJ...599..661L} for an assumed
$1-d$ flux-rope picture of the magnetic field configuration.

To summarise our results, we find that both the one and two dimensional
scenarios for the magnetic structures predict rotation measures which are of
the same order of magnitude as the observed ones. In cases where observational
estimates for field strength and length scales existed, there is also a better
than order of magnitude agreement. The explosive dynamo seems to underpredict
the observed Faraday rotation by a factor of two, due to the shorter magnetic
length scale predicted. In general, we would not expect our model to be more
accurate than to within a factor of two. This is especially true for our rough
parametrisation of the saturation state of the explosive dynamo.

\begin{table*}[t]
  \begin{tabular}{|l|cccccccccc|}
\hline
$ $cluster properties$           $  $   $ & Hydr. & Cent. & Cygn. & A1958 & A2597 & 3C31 & 
 
 Pers. & A85 & A2199 & Virgo \\
\hline
$ \mathrm{n}_{\mathrm{cc}} $  $ \;\;[\mathrm{10^{-3} \, /ccm}]          $ & 56.1 & 80.6 & 153 & 189 & 73.5 & 180 & 
 
 50.8 & 30.8 & 33.7 & 151 \\
$ \mathrm{T}_{\mathrm{cc}} $  $ \;\;[\mathrm{keV}]            $ & 2.7 & 2.2 & 6.5 & 3.7 & 1.3 & 0.7 & 3 & 5.5 & 1.6 & 1 \\
$ \mathrm{r}_{\mathrm{cc}} $  $ \;\;[\mathrm{kpc}]            $ & 35.5 & 8.57 & 10.7 & 13.4 & 28 & 1.2 & 57 & 
 
 45 & 29 & 1.6 \\
$ \mathrm{RM}_{\mathrm{rms}}^{\mathrm{obs}}  $  $ \;\;[\mathrm{m^{-2}}]   $ & 1350 & 660 & 1200 & 2900 & 1080 & 25 & -- & -- & -- & -- \\
$ L_{\mathrm{cluster}} $  $ \;\;[\mathrm{10^{44} \, erg/s}] $ & 4.61 & 0.32 & 1.79 & 6.22 & 6.89 & 
 
 0.077 & 11.8 & 8.16 & 3.28 & 0.14 \\
$ L(<r_{\mathrm{cooling}})/L_{\mathrm{cluster}}        $  $   $ & 0.52 & 0.41 & 0.49 & -- & 0.61 & -- & 0.59 & 0.3 & 0.47 & 0.5 \\
$ L_{\mathrm{cc}}/L_{\mathrm{cluster}}        $  $   $ & 0.37 & 0.14 & 0.3 & 0.2 & 0.14 & 
 
 0.0045 & 0.51 & 0.18 & 0.078 & 0.005 \\
\hline
hydrodynamical turbulence &
\multicolumn{10}{l|}{}\\
\hline
$ \lB \sim r_{\mathrm{bub}} $  $ \;\;[\mathrm{kpc}]           $ & 7.68 & 1.45 & 1.7 & 2.84 & 7.52 & 
 
 0.2 & 13.3 & 7.15 & 5.94 & 0.25 \\
$ \eps_{\mathrm{T}} $  $ \;\;[\mathrm{10^{-10} \, erg/ccm}] $ & 0.29 & 0.16 & 0.76 & 1.27 & 0.35 & 
 
 0.12 & 0.37 & 0.13 & 0.089 & 0.11 \\
$ v_{\mathrm{T}}  $  $ \;\;[\mathrm{km/s}]           $ & 250 & 156 & 244 & 283 & 240 & 87.5 & 
 
 295 & 225 & 177 & 94 \\
\hline
magnetic turbulence &
\multicolumn{10}{l|}{$d=1\,$ $\Rc=20$ -- {\it flux rope dynamo}}\\
\hline

$ \lambda_B $  $ \;\;[\mathrm{kpc}]        $ & 1.72 & 0.32 & 0.38 & 0.64 & 1.68 & 
 
 0.045 & 2.97 & 1.6 & 1.33 & 0.056 \\
$ B_\mathrm{rms} $   $ \;\;[\mathrm{\mu G}]              $ & 6.06 & 4.55 & 9.78 & 12.6 & 6.68 & 3.8 & 
 
 6.82 & 4.04 & 3.34 & 3.74 \\
$ \kappa_{\mathrm{visc}} $  $ \;\;[\mathrm{10^{27} \,  cm^2/s}]  $ & 7.89 & 0.93 & 1.71 & 3.31 & 7.43 & 
 
 0.073 & 16.1 & 6.61 & 4.33 & 0.097 \\
$ \mathrm{RM}_{\mathrm{rms}}^{\mathrm{exp}}  $  $ \;\;[\mathrm{m^{-2}}]   $ & 1520 & 351 & 1730 & 4010 & 1930 & 
 
 91.5 & 2590 & 607 & 400 & 96.9 \\
$ \mathrm{RM}_{\mathrm{rms}}^{\mathrm{obs}}/\mathrm{RM}_{\mathrm{rms}}^{\mathrm{exp}}$  $  $ & 0.89 & 1.88 & 0.69 & 0.72 & 0.56 & 
 
 0.27 & -- & -- & -- & -- \\
\hline
magnetic turbulence &
\multicolumn{10}{l|}{$d=2\,$ $\Rc=35$ -- {\it fluctuation dynamo}}\\
\hline

$ \lambda_B $  $ \;\;[\mathrm{kpc}]        $ & 1.3 & 0.25 & 0.29 & 0.48 & 1.27 & 
 
 0.034 & 2.25 & 1.21 & 1 & 0.042 \\
$ B_\mathrm{rms} $   $ \;\;[\mathrm{\mu G}]              $ & 11.1 & 8.36 & 18 & 23.2 & 12.3 & 6.99 & 
 
 12.5 & 7.44 & 6.13 & 6.88 \\
$ \mathrm{RM}_{\mathrm{rms}}^{\mathrm{exp}}  $  $ \;\;[\mathrm{m^{-2}}]   $ & 2430 & 561 & 2770 & 6420 & 3090 & 
 
 146 & 4140 & 970 & 640 & 155 \\
$ \mathrm{RM}_{\mathrm{rms}}^{\mathrm{obs}}/\mathrm{RM}_{\mathrm{rms}}^{\mathrm{exp}}$  $  $ & 0.55 & 1.18 & 0.43 & 0.45 & 0.35 & 
 
 0.17 & -- & -- & -- & -- \\
\hline
magnetic turbulence &
\multicolumn{10}{l|}{$d=2\,$ $\Rc=\Rc^*$ -- {\it explosive dynamo}}\\
\hline

$ \Rc     $  $             $ & 1060 & 461 & 461 & 922 & 1630 & 308 & 
 
 1380 & 541 & 839 & 279 \\
$ \lambda_B $  $ \;\;[\mathrm{kpc}]        $ & 0.24 & 0.068 & 0.079 & 0.094 & 0.19 & 
 
 0.012 & 0.36 & 0.31 & 0.2 & 0.015 \\
$ B_\mathrm{rms} $   $ \;\;[\mathrm{\mu G}]              $ & 4.75 & 4.39 & 9.44 & 10.3 & 4.7 & 4.06 & 
 
 5 & 3.75 & 2.77 & 4.09 \\
$ \mathrm{RM}_{\mathrm{rms}}^{\mathrm{exp}}  $  $ \;\;[\mathrm{m^{-2}}]   $ & 443 & 155 & 762 & 1250 & 452 & 
 
 49.3 & 659 & 247 & 131 & 54.9 \\
$ \mathrm{RM}_{\mathrm{rms}}^{\mathrm{obs}}/\mathrm{RM}_{\mathrm{rms}}^{\mathrm{exp}}$  $  $ & 3.04 & 4.27 & 1.57 & 2.32 & 2.39 & 
 
 0.51 & -- & -- & -- & -- \\
\hline
\end{tabular}

  \caption{Application of the model to cool cores of the Hydra~A (Hydr.),
  Centaurus (Cent.), Cygnus A (Cygn.), A1985, A2597, 3C31, Perseus (Pers.),
  A85, A2199 and Virgo clusters of galaxies. The first part of the table
  contains the cluster parameters whose values were taken from the
  literature. References and comments on possible biases of the data are given
  in Sect.~\ref{sec:application} and in detail in Appendix~\ref{app:A}. The
  second part describes the expected hydrodynamical turbulence. The third,
  fourth, and fifth parts describe the expected magnetic turbulence for the
  scenarios of magnetic flux ropes ($d=1$, $\Rc = 20$), flux sheets ($d=2$ $\Rc
  = 35$), and flux sheets in the explosive dynamo scenario ($d=2$ $\Rc =
  \Rc^*$), respectively. Numbers are given here to three digits accuracy,
  irrespective of the fact that the real uncertainties of most quantities are
  much larger.}
\label{tab:cc} 
\end{table*}

\section{Conclusion\label{sec:concl}}

We showed that many properties of magnetic field measurements in cool core
clusters, and especially in the case of Hydra~A, which we investigate in
detail, seem to support the picture that a saturated small-scale turbulent
dynamo is maintaining the magnetic fields. Although there exist dramatic
differences between small-scale dynamo scenarios, their saturated state might
be quite similar. If it can be assumed to be mostly a balance between the
turbulent entangling and Alfv\'enic straightening of magnetic field lines, the
state is characterised by only two numbers: the effective dimensionality of the
magnetic structures and the ratio between turbulence injection scale and
magnetic correlation length. Casting three representative small-scale dynamo
descriptions into this formulation, we find that they are all in rough
agreement with the data, which is as much as we can expect given the
approximate nature of our treatment. The investigated scenarios are the {\it
flux rope dynamo} \citep[e.g][ with a renormalised Reynolds
number]{1990alch.book.....Z, 1999PhRvL..83.2957S}, the {\it fluctuation dynamo}
as described in \citet{2006MNRAS.366.1437S}, and the {\it explosive dynamo}
introduced by \citet{2005ApJ...629..139S}.

The likely energy sources of the expected turbulence are buoyant radio bubbles
from the central galaxy, which can inject turbulence with the right amount of
power, and also on length scale, that fit the observed magnetic correlation
scales well.

Motivated by these indications of the physical processes in cool core clusters,
we developed a steady state scenario for the hydrodynamical and magnetic
turbulence, assuming that the turbulent feedback from the central radio source
compensates for the cool core radiative losses. This scenario predicts
turbulent length scales and energy densities that can be compared to
observations.

The hydrodynamical turbulence could have been probed with the
Astro-E2 satellite mission which unfortunately failed. We hope that
successor missions will permit the determination of the gas velocity
dispersion \citep{2003AstL...29..783S, 2003AstL...29..791I,
2005astro.ph..3656B}. At present, the magnetic turbulence can be
tested with the existing Faraday rotation measurements of extended
radio sources in cool core clusters. The Faraday dispersion is a
combined measure of the magnetic field strength and correlation
length.

For a sample of prominent cool core clusters, we calculate the expected hydro-
and magnetic-turbulence, and the predicted Faraday dispersion. In cases of
existing Faraday measurements, we find that our estimates reproduce the
observed magnitude of the dispersion over roughly two orders of magnitude in
\RM (or four in $\langle \RM^2 \rangle$). On average, our predictions for the
models based on the {\it flux rope} and the {\it fluctuating} dynamo
\citep{1999PhRvL..83.2957S, 2006MNRAS.366.1437S} are a factor of two higher
than the measurements, which is not too surprising, since the actual data
sample larger cluster radii, whereas our estimates are aimed at a radio source
at the cluster centre. The predictions for the Faraday dispersion in the {\it
explosive dynamo} scenario are a factor of two below the observation as shown
by \cite{2006astro.ph..1246S}, which is more severe, since geometrical
considerations will only enlarge this discrepancy. However, the description of
the saturated state of this dynamos is currently more a guess than a precise
estimate. Thus, also this model should be regarded as in agreement with the
data.

Very little fine tuning went into our model, but this is likely more a
coincidence than a proof of the investigated scenario. There are various places
in the calculation where factors of order one could have been introduced. One
example is the assumption of the bubble radius being exactly the turbulence
injection scale. This assumption will require verification by numerical
simulations of the hydrodynamics, which is beyond the scope of this initial
investigation.  Another uncertain area is the adopted parameters of the dynamo
theory, which are also not fully settled in the literature.

Nevertheless, we have shown that the straightforward application of current
concepts of turbulent magnetic dynamos in combination with the emerging picture
of cool core stabilisation via heat injection due to the dissipation of
turbulence seeded by radio galaxy feedback leads to expectations for the
Faraday rotation signal that match well with the observations. This is the case
for a variety of galaxy clusters spanning two orders of magnitude in their
X-ray luminosities. Therefore, our picture of cluster cool core heating by
radio galaxy feedback in combination with ideas about the properties of the
saturated state of non-helical, small-scale magnetic dynamos passed a critical
test. Our scenario provides a number of testable predictions, especially about
the level of fluid turbulence in individual clusters. We hope that future
measurements of the plasma properties in galaxy clusters will even permit
discrimination between the different small-scale dynamo models.

\begin{acknowledgements}
We acknowledge stimulating discussions with K. Subramanian and A. Schekochihin
on the theory of small-scale dynamos. We are thankful to R. Laing for providing
us with the results on 3C31 prior to publication, and to V. Springel and
E. Churazov for comments on the manuscript.
\end{acknowledgements}

\appendix
\section{Individual clusters\label{app:A}}

In this Appendix the observational data is discussed. This collection of
literature values for cool core cluster parameters made use of the work done by
\citet{2004A&A...413...17P}.

\subsection{Hydra~A cluster\label{sec:HAcl}}

The gas parameters for the Hydra~A cluster are taken from
\citet{1997ApJ...491...38M}, and the X-ray luminosity is provided by
\citet{1997ApJ...481..660I}. The first detailed \RM map of Hydra~A was
published by \citet{1993ApJ...416..554T}, which exhibited very large \RM
values, especially for the south lobe of Hydra~A. The data was reanalysed using
the PACERMAN algorithm \citep{2005MNRAS.358..726D, 2005MNRAS.358..732V}, which
removed all areas in the northern lobe, and many of the southern lobe with
extremely high \RM values, indicating that those were observational
artefacts. Since the southern lobe \RM data seem to be still contaminated by
observational artefacts, as statistical tests for anti-correlations of
gradients of polarisation angles and \RM values reveal
\citep{2003ApJ...597..870E}, we use only the dispersion of the northern lobe,
as reported by \citet{2005MNRAS.358..732V}. The same \RM map was analysed by
\citet{2005A&A...434...67V} using a maximum-likelihood power spectra estimator
(see Fig.~\ref{fig:BturbHydra}), revealing central magnetic field strength of
$B_\mathrm{rms} \approx 7 \,\mu$G with an autocorrelation length of
$\lambda_B \approx 3\, $kpc.  

We note that model expectations and observational values of Faraday
dispersion, magnetic field strength and autocorrelation length agree
well. Since the Hydra~A north lobe is known to approach the observer (and the
radio emission extend to larger radii), the observational \RM dispersion should
be slightly biased to lower values.

\subsection{Centaurus cluster\label{sec:Cecl}}

The central electron density and the cool core radius are taken from
\citet{1999ApJ...517..627M}\footnote{Using the extremely small core radius
given in \citet{2002MNRAS.334..769T} leads to a cool core fractional luminosity
of only 3\%, which is obviously much smaller than the 40\% observed
\citep{1998MNRAS.298..416P}.}, and the central temperature from
\citet{2000MNRAS.312..663W}\footnote{Lower central temperatures have been
reported by \citet{2002MNRAS.331..273S} and \citet{2002MNRAS.334..769T}, but
since they were not used for the electron density estimates, and since our
magnetic field and \RM predictions are quite insensitive to the temperature, we
do not use them.}.  The observed Faraday dispersion for the Centaurus cluster
is derived from the embedded radio source PKS~1246-410 by
\citet{2002MNRAS.334..769T}. The published Faraday rotation map reveals several
isolated patches of extreme \RM values, and the histogram of \RM values
exhibits multiple peaks. A statistical analysis of the alignment of \RM
gradients and polarisation angle gradients reveals a significant
anti-correlation between the two \citep{2003ApJ...597..870E}, which is also an
indication of observational artefacts in this map. Thus, the reported \RM
dispersion may be too large.  However, an order of magnitude agreement between
observed and expected \RM dispersion is found.

\citet{2002MNRAS.334..769T} report a \textit{RM}-correlation length of
$\lambda_{RM} \sim 0.7\,$kpc and a field strength of $9\, \mu$G, which is
significantly different to our expectations. The different length scale is not
surprising, since usually $\lambda_{RM}> \lB$ \citep{2003A&A...401..835E}. The
larger observationally derived magnetic field strength may be a result of the
very small core radius and/or the possible overestimate of the \RM dispersion
reported in \citet{2002MNRAS.334..769T}.

Thus, the X-ray and radio observational data of the cool core of the Centaurus
cluster would need further improvements before conclusive statements about an
agreement or disagreement of our model and observations can be made in this
case. Both the observed and the expected \RM dispersion could be subject to
re-adjustments. The complex morphology of this cool core probably causes the
difficulties singling out a {\it `real'} core radius and central electron
density.

\subsection{Abell 1958 cluster (3C295) \label{sec:A1985cl}}

The electron density, central temperature, cool core radius, and
dispersion measure are taken from \citet{2001MNRAS.324..842A} and the
cluster luminosity from \citet{2003MNRAS.342..287A}.

The agreement of predicted and observed Faraday dispersion is very
good. Furthermore, the rather high predicted magnetic field strength
of $10-20\,\mu$G is also found observationally, since
\citet{2001MNRAS.324..842A} report $14\,\mu$G.

\subsection{Cygnus A cluster\label{sec:CAcl}}

We use the electron density, temperature and cool core radius reported in
\citet{1999ApJ...517..627M}. We measure the \RM dispersion to be
$1200\,\mathrm{m^{-2}}$ from the map published by \citet{1987ApJ...316..611D}
after removing pixels from the map with extraordinary large \RM jumps, which we
associate with observational artefacts ($n\pi$-ambiguities). Similar to the
case of Hydra~A, we see an asymmetry in the \RM dispersion between the two
lobes, which is a consequence of the different depth of the lobes \citep[the
Laing-Garrington effect:][]{1988Natur.331..149L,
1988Natur.331..147G}. Unfortunately, the two radio lobes are located well
beyond the cool core radius, which is even more true for their polarised
regions used for \RM measurements. Since we expect a declining magnetic field
strength as a function of radius, the reported \RM dispersion is likely smaller
than the one that would be measured from a radio source located in the centre
of the cool core, as discussed in Sect.~\ref{sec:RMprofile} and illustrated in
Fig.~\ref{fig:RMprofile}.  The predicted central \RM dispersion is a factor of
two smaller than the peripherally measured one. With our current understanding
of the geometry, there is no apparent conflict between these numbers.

\subsection{Perseus cluster\label{sec:Pecl}}

We take the central electron density, temperature and the cool core
radius from \citet{2003ApJ...590..225C} and the cluster X-ray
luminosity from \citet{1993ApJ...412..479D}. 

The Perseus cluster cool core exhibits diffuse radio emission, a radio
mini-halo. One can apply the classical minimum-energy arguments to the radio
halo data to search for the minimum of the sum of the magnetic and relativistic
particle energy densities necessary to reproduce the radio synchrotron
emission, and assume a constant proton to electron ratio
$k_\mathrm{p}$. \citet{2004MNRAS.352...76P} report a classical minimum field
strength of $7.2_{-0.4}^{+4.5}\, \mu$G (assuming $k_\mathrm{p} =1$), where the
{\it confidence interval} is given by the requirement that the energy density
should be within one $e$-fold from the minimum. \citet{2004MNRAS.352...76P}
also develop and apply the hadronic minimum-energy criteria to the Perseus
mini-halo. This criterion assumes that the relativistic electrons are injected
by hadronic interactions of a relativistic proton population, which usually
dominates the relativistic energy budget. No proton-to-electron factor
$k_\mathrm{p}$ has to be assumed in this case, since the physics of the
hadronic interaction determines this ratio. Applied to the Perseus radio halo,
a very similar central field strength of $8.8_{-5.4}^{+13.8}\, \mu$G was
found. Our cool core model predicts a field strength of $7-13\,\mu$G, which is
in agreement with these findings. A significant lower central field value of
about $1-3\,\mu$G was reported by \citet{2005MNRAS.360..133S}, based on an
Inverse Compton interpretation of a hard photon component in the X-ray spectra
of the cool core region. If the hard photons flux is due to another physical
mechanism, the derived value will become a lower limit. If the Inverse Compton
nature of the flux could be verified, this will be a very good field estimate
in the case of a non-intermittent field distribution. If there is significant
magnetic intermittency, spatially inhomogeneous electron cooling can produce an
anti-correlation between fields and relativistic electrons, which can easily
lead to Inverse Compton based magnetic field estimates lower by a factor of two
\citep{1999AA...344..409E, 2004JKAS...37..439E}.

As stated before, \citet{2003MNRAS.344L..48F} argued for a large-scale
viscosity in the Perseus cluster cool core of at least $4\cdot
10^{27}\, {\rm cm^2/s}$, which is in good agreement with the $1.6\cdot
10^{28}\, {\rm cm^2/s}$ predicted in our $1-d$ scenario.
\subsection{Abell 2597}

This cool core galaxy cluster has been observed with Chandra by
\citet{2001ApJ...562L.149M}. The observation has been reanalysed
recently by \citet{2005MNRAS.359.1229P}. The gas parameters are taken
from this reanalysis. The total luminosity used is taken from
\citet{1993ApJ...412..479D}.

\citet{2005MNRAS.359.1229P} study also the polarisation properties of the
central radio source in this cluster. They determine a \RM dispersion of 1080
rad m$^{-2}$, comparable to our expectations. They conclude that the cluster
magnetic field has a minimum magnetic field strength of 2.1 $\mu$G. Our model
predicts a higher magnetic field strength of 7-12 $\mu$G which is in agreement
with the lower limit of \citet{2005MNRAS.359.1229P}.

\subsection{3C31}

The cool X-ray environment of the extended radio source 3C31 is associated with
a group of galaxies. X-ray measurements suggest that this group shows
properties similar to a cool core of a galaxy cluster. Therefore, it seems to
be a good case to test our model in a different situation. The parameters for
the gas were taken from \citet{2002MNRAS.334..182H}. However we use the total
luminosity from \citet{1999A&A...344..755K}.

Detailed \RM measurements have been carried out by Laing et al. (in
prep.). They determine the \RM dispersion to be 25 rad m$^{-2}$
(Laing, private communication).

\subsection{Abell 85}

We use the electron density, temperature, cool core radius, and luminosity
reported in \citet{1999ApJ...517..627M}, and \citet{1993ApJ...412..479D}. To
our knowledge, no \RM measurements or magnetic field estimates are available
for the centre of this and all following clusters. Thus, our $\RM$ results are
predictions.

\subsection{Abell 2199}

The parameters of the electron density profile as reported in
\citet{1999ApJ...517..627M} were used. The temperature of the cool core
are taken from \citet{2002MNRAS.335L...7V} and the total luminosity
from \citet{1993ApJ...412..479D}.

\subsection{Virgo cluster}

The Virgo cluster was observed with XMM-Newton by
\citet{2002A&A...386...77M}. The gas parameters are taken from their
analysis. However the total luminosity of \citet{1993ApJ...412..479D}
is used for our calculations. The large difference between the cool core
luminosity and the luminosity within the (larger) cooling radius is due to the
extremely shallow density profile ($\beta \approx 0.5$) of Virgo. However, this
should not affect our \RM estimates too much, since only a $\beta \approx 0.25$
should lead to problems, as was argued in Sect.~\ref{sec:RMprofile}.

\bibliography{aamnem99,../Bib/tae}

\begin{thebibliography}{141}
\expandafter\ifx\csname natexlab\endcsname\relax\def\natexlab#1{#1}\fi

\bibitem[{{Allen}(1995)}]{1995MNRAS.276..947A}
{Allen}, S.~W. 1995, \mnras, 276, 947

\bibitem[{{Allen} {et~al.}(2003){Allen}, {Schmidt}, {Fabian}, \&
  {Ebeling}}]{2003MNRAS.342..287A}
{Allen}, S.~W., {Schmidt}, R.~W., {Fabian}, A.~C., \& {Ebeling}, H. 2003,
  \mnras, 342, 287

\bibitem[{{Allen} {et~al.}(2001){Allen}, {Taylor}, {Nulsen}, {Johnstone},
  {David}, {Ettori}, {Fabian}, {Forman}, {Jones}, \&
  {McNamara}}]{2001MNRAS.324..842A}
{Allen}, S.~W., {Taylor}, G.~B., {Nulsen}, P.~E.~J., {et~al.} 2001, \mnras,
  324, 842

\bibitem[{{B{\" o}hringer} {et~al.}(2002){B{\" o}hringer}, {Matsushita},
  {Churazov}, {Ikebe}, \& {Chen}}]{2002A&A...382..804B}
{B{\" o}hringer}, H., {Matsushita}, K., {Churazov}, E., {Ikebe}, Y., \& {Chen},
  Y. 2002, \aap, 382, 804

\bibitem[{{Batchelor}(1950)}]{Batchelor1950}
{Batchelor}, G. 1950, Proc. R. Soc. London A, 201, 405

\bibitem[{{Bertone} {et~al.}(2005){Bertone}, {Vogt}, \&
  {En{\ss}lin}}]{bertone06}
{Bertone}, S., {Vogt}, C., \& {En{\ss}lin}, T.~A. 2005, \mnras submitted

\bibitem[{{Bicknell} {et~al.}(1990){Bicknell}, {Cameron}, \&
  {Gingold}}]{1990ApJ...357..373B}
{Bicknell}, G.~V., {Cameron}, R.~A., \& {Gingold}, R.~A. 1990, \apj, 357, 373

\bibitem[{{Br{\" u}ggen} \& {Kaiser}(2001)}]{2001MNRAS.325..676B}
{Br{\" u}ggen}, M. \& {Kaiser}, C.~R. 2001, \mnras, 325, 676

\bibitem[{{Br{\" u}ggen} {et~al.}(2002){Br{\" u}ggen}, {Kaiser}, {Churazov}, \&
  {En{\ss}lin}}]{2002MNRAS.331..545B}
{Br{\" u}ggen}, M., {Kaiser}, C.~R., {Churazov}, E., \& {En{\ss}lin}, T.~A.
  2002, \mnras, 331, 545

\bibitem[{{Brandenburg} \& {Subramanian}(2004)}]{2004astro.ph..5052B}
{Brandenburg}, A. \& {Subramanian}, K. 2004, ArXiv Astrophysics e-prints

\bibitem[{{Br{\"u}ggen} {et~al.}(2005{\natexlab{a}}){Br{\"u}ggen}, {Hoeft}, \&
  {Ruszkowski}}]{2005astro.ph..3656B}
{Br{\"u}ggen}, M., {Hoeft}, M., \& {Ruszkowski}, M. 2005{\natexlab{a}}, ArXiv
  Astrophysics e-prints

\bibitem[{{Br{\"u}ggen} {et~al.}(2005{\natexlab{b}}){Br{\"u}ggen},
  {Ruszkowski}, \& {Hallman}}]{2005ApJ...630..740B}
{Br{\"u}ggen}, M., {Ruszkowski}, M., \& {Hallman}, E. 2005{\natexlab{b}}, \apj,
  630, 740

\bibitem[{{Carilli} \& {Taylor}(2002)}]{2002ARA&A..40..319C}
{Carilli}, C.~L. \& {Taylor}, G.~B. 2002, \araa, 40, 319

\bibitem[{{Cen}(2005)}]{2005ApJ...620..191C}
{Cen}, R. 2005, \apj, 620, 191

\bibitem[{{Chakrabarti} {et~al.}(1994){Chakrabarti}, {Rosner}, \&
  {Vainshtein}}]{1994Natur.368..434C}
{Chakrabarti}, S.~K., {Rosner}, R., \& {Vainshtein}, S.~I. 1994, \nat, 368, 434

\bibitem[{{Chandran}(2004)}]{2004ApJ...616..169C}
{Chandran}, B.~D.~G. 2004, \apj, 616, 169

\bibitem[{{Chandran} \& {Maron}(2004)}]{2004ApJ...602..170C}
{Chandran}, B.~D.~G. \& {Maron}, J.~L. 2004, \apj, 602, 170

\bibitem[{{Cho} {et~al.}(2003){Cho}, {Lazarian}, {Honein}, {Knaepen},
  {Kassinos}, \& {Moin}}]{2003ApJ...589L..77C}
{Cho}, J., {Lazarian}, A., {Honein}, A., {et~al.} 2003, \apjl, 589, L77

\bibitem[{{Cho} \& {Vishniac}(2000)}]{2000ApJ...538..217C}
{Cho}, J. \& {Vishniac}, E.~T. 2000, \apj, 538, 217

\bibitem[{{Churazov} {et~al.}(2001){Churazov}, {Br{\" u}ggen}, {Kaiser}, {B{\"
  o}hringer}, \& {Forman}}]{2001ApJ...554..261C}
{Churazov}, E., {Br{\" u}ggen}, M., {Kaiser}, C.~R., {B{\" o}hringer}, H., \&
  {Forman}, W. 2001, \apj, 554, 261

\bibitem[{{Churazov} {et~al.}(2003){Churazov}, {Forman}, {Jones}, \& {B{\"
  o}hringer}}]{2003ApJ...590..225C}
{Churazov}, E., {Forman}, W., {Jones}, C., \& {B{\" o}hringer}, H. 2003, \apj,
  590, 225

\bibitem[{{Churazov} {et~al.}(2004){Churazov}, {Forman}, {Jones}, {Sunyaev}, \&
  {B{\" o}hringer}}]{2004MNRAS.347...29C}
{Churazov}, E., {Forman}, W., {Jones}, C., {Sunyaev}, R., \& {B{\" o}hringer},
  H. 2004, \mnras, 347, 29

\bibitem[{{Churazov} {et~al.}(2002){Churazov}, {Sunyaev}, {Forman}, \& {B{\"
  o}hringer}}]{2002MNRAS.332..729C}
{Churazov}, E., {Sunyaev}, R., {Forman}, W., \& {B{\" o}hringer}, H. 2002,
  \mnras, 332, 729

\bibitem[{{Clarke}(2004)}]{2004JKAS...37..337C}
{Clarke}, T.~E. 2004, Journal of Korean Astronomical Society, 37, 337

\bibitem[{{Clarke} \& {En{\ss}lin}(2005)}]{ClarkeEnsslin06}
{Clarke}, T.~E. \& {En{\ss}lin}, T.~A. 2005, {\aj} submitted

\bibitem[{{Clarke} {et~al.}(2001){Clarke}, {Kronberg}, \& {B{\"
  o}hringer}}]{2001ApJ...547L.111C}
{Clarke}, T.~E., {Kronberg}, P.~P., \& {B{\" o}hringer}, H. 2001, \apjl, 547,
  L111

\bibitem[{{Crawford} {et~al.}(1999){Crawford}, {Allen}, {Ebeling}, {Edge}, \&
  {Fabian}}]{1999MNRAS.306..857C}
{Crawford}, C.~S., {Allen}, S.~W., {Ebeling}, H., {Edge}, A.~C., \& {Fabian},
  A.~C. 1999, \mnras, 306, 857

\bibitem[{{Crawford} {et~al.}(2005){Crawford}, {Hatch}, {Fabian}, \&
  {Sanders}}]{2005MNRAS.363..216C}
{Crawford}, C.~S., {Hatch}, N.~A., {Fabian}, A.~C., \& {Sanders}, J.~S. 2005,
  \mnras, 363, 216

\bibitem[{{Daly} \& {Loeb}(1990)}]{1990ApJ...364..451D}
{Daly}, R.~A. \& {Loeb}, A. 1990, \apj, 364, 451

\bibitem[{{David} {et~al.}(1993){David}, {Slyz}, {Jones}, {Forman}, {Vrtilek},
  \& {Arnaud}}]{1993ApJ...412..479D}
{David}, L.~P., {Slyz}, A., {Jones}, C., {et~al.} 1993, \apj, 412, 479

\bibitem[{{De Young}(1992)}]{1992ApJ...386..464D}
{De Young}, D.~S. 1992, \apj, 386, 464

\bibitem[{{Dennis} \& {Chandran}(2005)}]{2005ApJ...622..205D}
{Dennis}, T.~J. \& {Chandran}, B.~D.~G. 2005, \apj, 622, 205

\bibitem[{{Dolag} {et~al.}(1999){Dolag}, {Bartelmann}, \&
  {Lesch}}]{1999A&A...348..351D}
{Dolag}, K., {Bartelmann}, M., \& {Lesch}, H. 1999, \aap, 348, 351

\bibitem[{{Dolag} {et~al.}(2002){Dolag}, {Bartelmann}, \&
  {Lesch}}]{2002A&A...387..383D}
{Dolag}, K., {Bartelmann}, M., \& {Lesch}, H. 2002, \aap, 387, 383

\bibitem[{{Dolag} {et~al.}(2004){Dolag}, {Jubelgas}, {Springel}, {Borgani}, \&
  {Rasia}}]{2004ApJ...606L..97D}
{Dolag}, K., {Jubelgas}, M., {Springel}, V., {Borgani}, S., \& {Rasia}, E.
  2004, \apjl, 606, L97

\bibitem[{{Dolag} {et~al.}(2001){Dolag}, {Schindler}, {Govoni}, \&
  {Feretti}}]{2001A&A...378..777D}
{Dolag}, K., {Schindler}, S., {Govoni}, F., \& {Feretti}, L. 2001, \aap, 378,
  777

\bibitem[{{Dolag} {et~al.}(2005){Dolag}, {Vogt}, \&
  {En{\ss}lin}}]{2005MNRAS.358..726D}
{Dolag}, K., {Vogt}, C., \& {En{\ss}lin}, T.~A. 2005, \mnras, 358, 726

\bibitem[{{Donahue} {et~al.}(2000){Donahue}, {Mack}, {Voit}, {Sparks},
  {Elston}, \& {Maloney}}]{2000ApJ...545..670D}
{Donahue}, M., {Mack}, J., {Voit}, G.~M., {et~al.} 2000, \apj, 545, 670

\bibitem[{{Dreher} {et~al.}(1987){Dreher}, {Carilli}, \&
  {Perley}}]{1987ApJ...316..611D}
{Dreher}, J.~W., {Carilli}, C.~L., \& {Perley}, R.~A. 1987, \apj, 316, 611

\bibitem[{{Edge}(2001)}]{2001MNRAS.328..762E}
{Edge}, A.~C. 2001, \mnras, 328, 762

\bibitem[{{Edge} \& {Frayer}(2003)}]{2003ApJ...594L..13E}
{Edge}, A.~C. \& {Frayer}, D.~T. 2003, \apjl, 594, L13

\bibitem[{{Eilek} \& {Owen}(2002)}]{2002ApJ...567..202E}
{Eilek}, J.~A. \& {Owen}, F.~N. 2002, \apj, 567, 202

\bibitem[{{En{\ss}lin}(2004)}]{2004JKAS...37..439E}
{En{\ss}lin}, T. 2004, Journal of Korean Astronomical Society, 37, 439

\bibitem[{{En{\ss}lin} {et~al.}(1998{\natexlab{a}}){En{\ss}lin}, {Biermann},
  {Klein}, \& {Kohle}}]{1998AA...332..395E}
{En{\ss}lin}, T.~A., {Biermann}, P.~L., {Klein}, U., \& {Kohle}, S.
  1998{\natexlab{a}}, \aap, 332, 395

\bibitem[{{En{\ss}lin} {et~al.}(1997){En{\ss}lin}, {Biermann}, {Kronberg}, \&
  {Wu}}]{1997ApJ...477..560E}
{En{\ss}lin}, T.~A., {Biermann}, P.~L., {Kronberg}, P.~P., \& {Wu}, X.-P. 1997,
  \apj, 477, 560

\bibitem[{{En{\ss}lin} \& {Heinz}(2002)}]{2002A&A...384L..27E}
{En{\ss}lin}, T.~A. \& {Heinz}, S. 2002, \aap, 384, L27

\bibitem[{{En{\ss}lin} {et~al.}(1999){En{\ss}lin}, {Lieu}, \&
  {Biermann}}]{1999AA...344..409E}
{En{\ss}lin}, T.~A., {Lieu}, R., \& {Biermann}, P.~L. 1999, \aap, 344, 409

\bibitem[{{En{\ss}lin} \& {Vogt}(2003)}]{2003A&A...401..835E}
{En{\ss}lin}, T.~A. \& {Vogt}, C. 2003, \aap, 401, 835

\bibitem[{{En{\ss}lin} {et~al.}(2003){En{\ss}lin}, {Vogt}, {Clarke}, \&
  {Taylor}}]{2003ApJ...597..870E}
{En{\ss}lin}, T.~A., {Vogt}, C., {Clarke}, T.~E., \& {Taylor}, G.~B. 2003,
  \apj, 597, 870

\bibitem[{{En{\ss}lin} {et~al.}(1998{\natexlab{b}}){En{\ss}lin}, {Wang},
  {Nath}, \& {Biermann}}]{1998AA...333L..47E}
{En{\ss}lin}, T.~A., {Wang}, Y., {Nath}, B.~B., \& {Biermann}, P.~L.
  1998{\natexlab{b}}, \aap, 333, L47

\bibitem[{{Fabian} {et~al.}(1991){Fabian}, {Nulsen}, \&
  {Canizares}}]{1991A&ARv...2..191F}
{Fabian}, A.~C., {Nulsen}, P.~E.~J., \& {Canizares}, C.~R. 1991, \aapr, 2, 191

\bibitem[{{Fabian} {et~al.}(2003{\natexlab{a}}){Fabian}, {Sanders}, {Allen},
  {Crawford}, {Iwasawa}, {Johnstone}, {Schmidt}, \&
  {Taylor}}]{2003MNRAS.344L..43F}
{Fabian}, A.~C., {Sanders}, J.~S., {Allen}, S.~W., {et~al.} 2003{\natexlab{a}},
  \mnras, 344, L43

\bibitem[{{Fabian} {et~al.}(2003{\natexlab{b}}){Fabian}, {Sanders}, {Crawford},
  {Conselice}, {Gallagher}, \& {Wyse}}]{2003MNRAS.344L..48F}
{Fabian}, A.~C., {Sanders}, J.~S., {Crawford}, C.~S., {et~al.}
  2003{\natexlab{b}}, \mnras, 344, L48

\bibitem[{{Feretti} {et~al.}(1995){Feretti}, {Dallacasa}, {Giovannini}, \&
  {Tagliani}}]{1995A&A...302..680F}
{Feretti}, L., {Dallacasa}, D., {Giovannini}, G., \& {Tagliani}, A. 1995, \aap,
  302, 680

\bibitem[{{Feretti} {et~al.}(1999){Feretti}, {Dallacasa}, {Govoni},
  {Giovannini}, {Taylor}, \& {Klein}}]{1999A&A...344..472F}
{Feretti}, L., {Dallacasa}, D., {Govoni}, F., {et~al.} 1999, \aap, 344, 472

\bibitem[{{Fujita} {et~al.}(2004){Fujita}, {Suzuki}, \&
  {Wada}}]{2004ApJ...600..650F}
{Fujita}, Y., {Suzuki}, T.~K., \& {Wada}, K. 2004, \apj, 600, 650

\bibitem[{{Garrington} {et~al.}(1988){Garrington}, {Leahy}, {Conway}, \&
  {Laing}}]{1988Natur.331..147G}
{Garrington}, S.~T., {Leahy}, J.~P., {Conway}, R.~G., \& {Laing}, R.~A. 1988,
  \nat, 331, 147

\bibitem[{{Ge} \& {Owen}(1993)}]{1993AJ....105..778G}
{Ge}, J.~P. \& {Owen}, F.~N. 1993, \aj, 105, 778

\bibitem[{{Goldman} \& {Rephaeli}(1991)}]{1991ApJ...380..344G}
{Goldman}, I. \& {Rephaeli}, Y. 1991, \apj, 380, 344

\bibitem[{{Goldreich} \& {Sridhar}(1997)}]{1997ApJ...485..680G}
{Goldreich}, P. \& {Sridhar}, S. 1997, \apj, 485, 680

\bibitem[{{Goldshmidt} \& {Rephaeli}(1993)}]{1993ApJ...411..518G}
{Goldshmidt}, O. \& {Rephaeli}, Y. 1993, \apj, 411, 518

\bibitem[{{Govoni} \& {Feretti}(2004)}]{2004IJMPD..13.1549G}
{Govoni}, F. \& {Feretti}, L. 2004, International Journal of Modern Physics D,
  13, 1549

\bibitem[{{Govoni} {et~al.}(2001){Govoni}, {Taylor}, {Dallacasa}, {Feretti}, \&
  {Giovannini}}]{2001A&A...379..807G}
{Govoni}, F., {Taylor}, G.~B., {Dallacasa}, D., {Feretti}, L., \& {Giovannini},
  G. 2001, \aap, 379, 807

\bibitem[{{Hansen} {et~al.}(1995){Hansen}, {Jorgensen}, \&
  {Norgaard-Nielsen}}]{1995A&A...297...13H}
{Hansen}, L., {Jorgensen}, H.~E., \& {Norgaard-Nielsen}, H.~U. 1995, \aap, 297,
  13

\bibitem[{{Hardcastle} {et~al.}(2002){Hardcastle}, {Worrall}, {Birkinshaw},
  {Laing}, \& {Bridle}}]{2002MNRAS.334..182H}
{Hardcastle}, M.~J., {Worrall}, D.~M., {Birkinshaw}, M., {Laing}, R.~A., \&
  {Bridle}, A.~H. 2002, \mnras, 334, 182

\bibitem[{{Haugen} {et~al.}(2004){Haugen}, {Brandenburg}, \&
  {Dobler}}]{2004PhRvE..70a6308H}
{Haugen}, N.~E., {Brandenburg}, A., \& {Dobler}, W. 2004, \pre, 70, 016308

\bibitem[{{Heckman} {et~al.}(1989){Heckman}, {Baum}, {van Breugel}, \&
  {McCarthy}}]{1989ApJ...338...48H}
{Heckman}, T.~M., {Baum}, S.~A., {van Breugel}, W.~J.~M., \& {McCarthy}, P.
  1989, \apj, 338, 48

\bibitem[{{Heinz} {et~al.}(1998){Heinz}, {Reynolds}, \&
  {Begelman}}]{1998ApJ...501..126H}
{Heinz}, S., {Reynolds}, C.~S., \& {Begelman}, M.~C. 1998, \apj, 501, 126

\bibitem[{{Hoeft} \& {Br{\" u}ggen}(2004)}]{2004ApJ...617..896H}
{Hoeft}, M. \& {Br{\" u}ggen}, M. 2004, \apj, 617, 896

\bibitem[{{Ikebe} {et~al.}(1997){Ikebe}, {Makishima}, {Ezawa}, {Fukazawa},
  {Hirayama}, {Honda}, {Ishisaki}, {Kikuchi}, {Kubo}, {Murakami}, {Ohashi},
  {Takahashi}, \& {Yamashita}}]{1997ApJ...481..660I}
{Ikebe}, Y., {Makishima}, K., {Ezawa}, H., {et~al.} 1997, \apj, 481, 660

\bibitem[{{Inogamov} \& {Sunyaev}(2003)}]{2003AstL...29..791I}
{Inogamov}, N.~A. \& {Sunyaev}, R.~A. 2003, Astronomy Letters, 29, 791

\bibitem[{{Jaffe}(1980)}]{1980ApJ...241..925J}
{Jaffe}, W. 1980, \apj, 241, 925

\bibitem[{{Jaffe} \& {Bremer}(1997)}]{1997MNRAS.284L...1J}
{Jaffe}, W. \& {Bremer}, M.~N. 1997, \mnras, 284, L1

\bibitem[{{Johnston-Hollitt} \& {Ekers}(2004)}]{2004astro.ph.11045J}
{Johnston-Hollitt}, M. \& {Ekers}, R.~D. 2004, ArXiv:astro-ph/0411045

\bibitem[{{Jubelgas} {et~al.}(2004){Jubelgas}, {Springel}, \&
  {Dolag}}]{2004MNRAS.351..423J}
{Jubelgas}, M., {Springel}, V., \& {Dolag}, K. 2004, \mnras, 351, 423

\bibitem[{{Kaiser} {et~al.}(2005){Kaiser}, {Pavlovski}, {Pope}, \&
  {Fangohr}}]{2005MNRAS.359..493K}
{Kaiser}, C.~R., {Pavlovski}, G., {Pope}, E.~C.~D., \& {Fangohr}, H. 2005,
  \mnras, 359, 493

\bibitem[{{Kazantsev}(1967)}]{kazantsev67}
{Kazantsev}, A. 1967, J. Exp. Theor. Phys., 53, 1806

\bibitem[{{Kim} {et~al.}(1991){Kim}, {Kronberg}, \&
  {Tribble}}]{1991ApJ...379...80K}
{Kim}, K.-T., {Kronberg}, P.~P., \& {Tribble}, P.~C. 1991, \apj, 379, 80

\bibitem[{{Komossa} \& {B\"ohringer}(1999)}]{1999A&A...344..755K}
{Komossa}, S. \& {B\"ohringer}, H. 1999, \aap, 344, 755

\bibitem[{{Kronberg} {et~al.}(2001){Kronberg}, {Dufton}, {Li}, \&
  {Colgate}}]{2001ApJ...560..178K}
{Kronberg}, P.~P., {Dufton}, Q.~W., {Li}, H., \& {Colgate}, S.~A. 2001, \apj,
  560, 178

\bibitem[{{Kronberg} {et~al.}(1999){Kronberg}, {Lesch}, \&
  {Hopp}}]{1999ApJ...511...56K}
{Kronberg}, P.~P., {Lesch}, H., \& {Hopp}, U. 1999, \apj, 511, 56

\bibitem[{{Laing}(1988)}]{1988Natur.331..149L}
{Laing}, R.~A. 1988, \nat, 331, 149

\bibitem[{{Loewenstein} \& {Fabian}(1990)}]{1990MNRAS.242..120L}
{Loewenstein}, M. \& {Fabian}, A.~C. 1990, \mnras, 242, 120

\bibitem[{{Longcope} {et~al.}(2003){Longcope}, {McLeish}, \&
  {Fisher}}]{2003ApJ...599..661L}
{Longcope}, D.~W., {McLeish}, T.~C.~B., \& {Fisher}, G.~H. 2003, \apj, 599, 661

\bibitem[{{Malyshkin}(2001)}]{2001ApJ...554..561M}
{Malyshkin}, L. 2001, \apj, 554, 561

\bibitem[{{Matsushita} {et~al.}(2002){Matsushita}, {Belsole}, {Finoguenov}, \&
  {B{\" o}hringer}}]{2002A&A...386...77M}
{Matsushita}, K., {Belsole}, E., {Finoguenov}, A., \& {B{\" o}hringer}, H.
  2002, \aap, 386, 77

\bibitem[{{Mazzotta} {et~al.}(2004){Mazzotta}, {Brunetti}, {Giacintucci},
  {Venturi}, \& {Bardelli}}]{2004JKAS...37..381M}
{Mazzotta}, P., {Brunetti}, G., {Giacintucci}, S., {Venturi}, T., \&
  {Bardelli}, S. 2004, Journal of Korean Astronomical Society, 37, 381

\bibitem[{{McNamara} {et~al.}(2001){McNamara}, {Wise}, {Nulsen}, {David},
  {Carilli}, {Sarazin}, {O'Dea}, {Houck}, {Donahue}, {Baum}, {Voit},
  {O'Connell}, \& {Koekemoer}}]{2001ApJ...562L.149M}
{McNamara}, B.~R., {Wise}, M.~W., {Nulsen}, P.~E.~J., {et~al.} 2001, \apjl,
  562, L149

\bibitem[{{Mohr} \& {Evrard}(1997)}]{1997ApJ...491...38M}
{Mohr}, J.~J. \& {Evrard}, A.~E. 1997, \apj, 491, 38

\bibitem[{{Mohr} {et~al.}(1999){Mohr}, {Mathiesen}, \&
  {Evrard}}]{1999ApJ...517..627M}
{Mohr}, J.~J., {Mathiesen}, B., \& {Evrard}, A.~E. 1999, \apj, 517, 627

\bibitem[{{Narayan} \& {Medvedev}(2001)}]{2001ApJ...562L.129N}
{Narayan}, R. \& {Medvedev}, M.~V. 2001, \apjl, 562, L129

\bibitem[{{Nipoti} \& {Binney}(2004)}]{2004MNRAS.349.1509N}
{Nipoti}, C. \& {Binney}, J. 2004, \mnras, 349, 1509

\bibitem[{{O'Dea} {et~al.}(1998){O'Dea}, {Payne}, \&
  {Kocevski}}]{1998AJ....116..623O}
{O'Dea}, C.~P., {Payne}, H.~E., \& {Kocevski}, D. 1998, \aj, 116, 623

\bibitem[{{Oegerle} {et~al.}(2001){Oegerle}, {Cowie}, {Davidsen}, {Hu},
  {Hutchings}, {Murphy}, {Sembach}, \& {Woodgate}}]{2001ApJ...560..187O}
{Oegerle}, W.~R., {Cowie}, L., {Davidsen}, A., {et~al.} 2001, \apj, 560, 187

\bibitem[{{Peres} {et~al.}(1998){Peres}, {Fabian}, {Edge}, {Allen},
  {Johnstone}, \& {White}}]{1998MNRAS.298..416P}
{Peres}, C.~B., {Fabian}, A.~C., {Edge}, A.~C., {et~al.} 1998, \mnras, 298, 416

\bibitem[{{Perley} \& {Taylor}(1991)}]{1991AJ....101.1623P}
{Perley}, R.~A. \& {Taylor}, G.~B. 1991, \aj, 101, 1623

\bibitem[{{Pfrommer} \& {En{\ss}lin}(2004{\natexlab{a}})}]{2004A&A...413...17P}
{Pfrommer}, C. \& {En{\ss}lin}, T.~A. 2004{\natexlab{a}}, \aap, 413, 17

\bibitem[{{Pfrommer} \& {En{\ss}lin}(2004{\natexlab{b}})}]{2004MNRAS.352...76P}
{Pfrommer}, C. \& {En{\ss}lin}, T.~A. 2004{\natexlab{b}}, \mnras, 352, 76

\bibitem[{{Pollack} {et~al.}(2005){Pollack}, {Taylor}, \&
  {Allen}}]{2005MNRAS.359.1229P}
{Pollack}, L.~K., {Taylor}, G.~B., \& {Allen}, S.~W. 2005, \mnras, 359, 1229

\bibitem[{{Pringle}(1989)}]{1989MNRAS.239..479P}
{Pringle}, J.~E. 1989, \mnras, 239, 479

\bibitem[{{Quilis} {et~al.}(2001){Quilis}, {Bower}, \&
  {Balogh}}]{2001MNRAS.328.1091Q}
{Quilis}, V., {Bower}, R.~G., \& {Balogh}, M.~L. 2001, \mnras, 328, 1091

\bibitem[{{Rees}(1987)}]{1987QJRAS..28..197R}
{Rees}, M.~J. 1987, \qjras, 28, 197

\bibitem[{{Reynolds} {et~al.}(2002){Reynolds}, {Heinz}, \&
  {Begelman}}]{2002MNRAS.332..271R}
{Reynolds}, C.~S., {Heinz}, S., \& {Begelman}, M.~C. 2002, \mnras, 332, 271

\bibitem[{{Reynolds} {et~al.}(2005){Reynolds}, {McKernan}, {Fabian}, {Stone},
  \& {Vernaleo}}]{2005MNRAS.357..242R}
{Reynolds}, C.~S., {McKernan}, B., {Fabian}, A.~C., {Stone}, J.~M., \&
  {Vernaleo}, J.~C. 2005, \mnras, 357, 242

\bibitem[{{Roettiger} {et~al.}(1999{\natexlab{a}}){Roettiger}, {Burns}, \&
  {Stone}}]{1999ApJ...518..603R}
{Roettiger}, K., {Burns}, J.~O., \& {Stone}, J.~M. 1999{\natexlab{a}}, \apj,
  518, 603

\bibitem[{{Roettiger} {et~al.}(1999{\natexlab{b}}){Roettiger}, {Stone}, \&
  {Burns}}]{1999ApJ...518..594R}
{Roettiger}, K., {Stone}, J.~M., \& {Burns}, J.~O. 1999{\natexlab{b}}, \apj,
  518, 594

\bibitem[{{Roland}(1981)}]{1981A&A....93..407R}
{Roland}, J. 1981, \aap, 93, 407

\bibitem[{{Rudnick} \& {Blundell}(2003)}]{2003ApJ...588..143R}
{Rudnick}, L. \& {Blundell}, K.~M. 2003, \apj, 588, 143

\bibitem[{{Ruszkowski} \& {Begelman}(2002)}]{2002ApJ...581..223R}
{Ruszkowski}, M. \& {Begelman}, M.~C. 2002, \apj, 581, 223

\bibitem[{{Ruszkowski} {et~al.}(2004){Ruszkowski}, {Br{\" u}ggen}, \&
  {Begelman}}]{2004ApJ...615..675R}
{Ruszkowski}, M., {Br{\" u}ggen}, M., \& {Begelman}, M.~C. 2004, \apj, 615, 675

\bibitem[{{Ruzmaikin} {et~al.}(1989){Ruzmaikin}, {Sokolov}, \&
  {Shukurov}}]{1989MNRAS.241....1R}
{Ruzmaikin}, A., {Sokolov}, D., \& {Shukurov}, A. 1989, \mnras, 241, 1

\bibitem[{{Salom{\' e}} \& {Combes}(2003)}]{2003A&A...412..657S}
{Salom{\' e}}, P. \& {Combes}, F. 2003, \aap, 412, 657

\bibitem[{{Sanders} \& {Fabian}(2002)}]{2002MNRAS.331..273S}
{Sanders}, J.~S. \& {Fabian}, A.~C. 2002, \mnras, 331, 273

\bibitem[{{Sanders} {et~al.}(2005){Sanders}, {Fabian}, \&
  {Dunn}}]{2005MNRAS.360..133S}
{Sanders}, J.~S., {Fabian}, A.~C., \& {Dunn}, R.~J.~H. 2005, \mnras, 360, 133

\bibitem[{{Schekochihin} {et~al.}(2002){Schekochihin}, {Cowley}, {Maron}, \&
  {Malyshkin}}]{2002PhRvE..65a6305S}
{Schekochihin}, A., {Cowley}, S., {Maron}, J., \& {Malyshkin}, L. 2002, \pre,
  65, 016305

\bibitem[{{Schekochihin} \& {Cowley}(2006)}]{2006astro.ph..1246S}
{Schekochihin}, A.~A. \& {Cowley}, S.~C. 2006, ArXiv Astrophysics e-prints

\bibitem[{{Schekochihin} {et~al.}(2005){Schekochihin}, {Cowley}, {Kulsrud},
  {Hammett}, \& {Sharma}}]{2005ApJ...629..139S}
{Schekochihin}, A.~A., {Cowley}, S.~C., {Kulsrud}, R.~M., {Hammett}, G.~W., \&
  {Sharma}, P. 2005, \apj, 629, 139

\bibitem[{{Schuecker} {et~al.}(2004){Schuecker}, {Finoguenov}, {Miniati}, {B{\"
  o}hringer}, \& {Briel}}]{2004A&A...426..387S}
{Schuecker}, P., {Finoguenov}, A., {Miniati}, F., {B{\" o}hringer}, H., \&
  {Briel}, U.~G. 2004, \aap, 426, 387

\bibitem[{{Smith} {et~al.}(1997){Smith}, {Bohlin}, {Bothun}, {O'Connell},
  {Roberts}, {Neff}, {Smith}, \& {Stecher}}]{1997ApJ...478..516S}
{Smith}, E.~P., {Bohlin}, R.~C., {Bothun}, G.~D., {et~al.} 1997, \apj, 478, 516

\bibitem[{{Soker}(2003)}]{2003MNRAS.342..463S}
{Soker}, N. 2003, \mnras, 342, 463

\bibitem[{{Soker}(2004)}]{2004MNRAS.350.1015S}
{Soker}, N. 2004, \mnras, 350, 1015

\bibitem[{{Soker} {et~al.}(2004){Soker}, {Blanton}, \&
  {Sarazin}}]{2004A&A...422..445S}
{Soker}, N., {Blanton}, E.~L., \& {Sarazin}, C.~L. 2004, \aap, 422, 445

\bibitem[{{Soker} \& {Pizzolato}(2005)}]{2005ApJ...622..847S}
{Soker}, N. \& {Pizzolato}, F. 2005, \apj, 622, 847

\bibitem[{{Soker} \& {Sarazin}(1990)}]{1990ApJ...348...73S}
{Soker}, N. \& {Sarazin}, C.~L. 1990, \apj, 348, 73

\bibitem[{{Sokolov} {et~al.}(1990){Sokolov}, {Ruzmaikin}, \&
  {Shukurov}}]{1990IAUS..140..499S}
{Sokolov}, D.~D., {Ruzmaikin}, A.~A., \& {Shukurov}, A. 1990, in IAU Symp. 140:
  Galactic and Intergalactic Magnetic Fields, 499--502

\bibitem[{{Subramanian}(1999)}]{1999PhRvL..83.2957S}
{Subramanian}, K. 1999, Physical Review Letters, 83, 2957

\bibitem[{{Subramanian} {et~al.}(2006){Subramanian}, {Shukurov}, \&
  {Haugen}}]{2006MNRAS.366.1437S}
{Subramanian}, K., {Shukurov}, A., \& {Haugen}, N.~E.~L. 2006, \mnras, 366,
  1437

\bibitem[{{Sunyaev} {et~al.}(2003){Sunyaev}, {Norman}, \&
  {Bryan}}]{2003AstL...29..783S}
{Sunyaev}, R.~A., {Norman}, M.~L., \& {Bryan}, G.~L. 2003, Astronomy Letters,
  29, 783

\bibitem[{{Taylor} {et~al.}(2002){Taylor}, {Fabian}, \&
  {Allen}}]{2002MNRAS.334..769T}
{Taylor}, G.~B., {Fabian}, A.~C., \& {Allen}, S.~W. 2002, \mnras, 334, 769

\bibitem[{{Taylor} {et~al.}(2001){Taylor}, {Govoni}, {Allen}, \&
  {Fabian}}]{2001MNRAS.326....2T}
{Taylor}, G.~B., {Govoni}, F., {Allen}, S.~W., \& {Fabian}, A.~C. 2001, \mnras,
  326, 2

\bibitem[{{Taylor} \& {Perley}(1993)}]{1993ApJ...416..554T}
{Taylor}, G.~B. \& {Perley}, R.~A. 1993, \apj, 416, 554

\bibitem[{{Tribble}(1993)}]{1993MNRAS.263...31T}
{Tribble}, P.~C. 1993, \mnras, 263, 31

\bibitem[{{Vogt} {et~al.}(2005){Vogt}, {Dolag}, \&
  {En{\ss}lin}}]{2005MNRAS.358..732V}
{Vogt}, C., {Dolag}, K., \& {En{\ss}lin}, T.~A. 2005, \mnras, 358, 732

\bibitem[{{Vogt} \& {En{\ss}lin}(2003)}]{2003A&A...412..373V}
{Vogt}, C. \& {En{\ss}lin}, T.~A. 2003, \aap, 412, 373

\bibitem[{{Vogt} \& {En{\ss}lin}(2005)}]{2005A&A...434...67V}
{Vogt}, C. \& {En{\ss}lin}, T.~A. 2005, \aap, 434, 67

\bibitem[{{Voigt} \& {Fabian}(2004)}]{2004MNRAS.347.1130V}
{Voigt}, L.~M. \& {Fabian}, A.~C. 2004, \mnras, 347, 1130

\bibitem[{{Voigt} {et~al.}(2002){Voigt}, {Schmidt}, {Fabian}, {Allen}, \&
  {Johnstone}}]{2002MNRAS.335L...7V}
{Voigt}, L.~M., {Schmidt}, R.~W., {Fabian}, A.~C., {Allen}, S.~W., \&
  {Johnstone}, R.~M. 2002, \mnras, 335, L7

\bibitem[{{V{\"o}lk} \& {Atoyan}(2000)}]{2000ApJ...541...88V}
{V{\"o}lk}, H.~J. \& {Atoyan}, A.~M. 2000, \apj, 541, 88

\bibitem[{{White}(2000)}]{2000MNRAS.312..663W}
{White}, D.~A. 2000, \mnras, 312, 663

\bibitem[{{Widrow}(2002)}]{2002RvMP...74..775W}
{Widrow}, L.~M. 2002, Reviews of Modern Physics, 74, 775

\bibitem[{{Zeldovich} {et~al.}(1990){Zeldovich}, {Ruzmaikin}, \&
  {Sokoloff}}]{1990alch.book.....Z}
{Zeldovich}, Y.~B., {Ruzmaikin}, A.~A., \& {Sokoloff}, D.~D. 1990, {The
  almighty chance} (World Scientific Lecture Notes in Physics, Singapore: World
  Scientific Publication, 1990)

\end{thebibliography}
\bibliographystyle{aa}

\end{document}